\documentclass[11pt]{article}
\usepackage[letterpaper,margin=1in]{geometry}
\usepackage[numbers,sort&compress]{natbib}
\usepackage{titletoc}
\usepackage{titlesec}
\usepackage[section]{placeins}
\usepackage{float}
\usepackage{needspace}
\usepackage[utf8]{inputenc}
\usepackage[T1]{fontenc}
\usepackage{hyperref}
\hypersetup{colorlinks=true,linkcolor=blue,urlcolor=blue,citecolor=blue}
\usepackage{xurl}
\usepackage{booktabs}
\usepackage{longtable}
\usepackage{multirow}
\usepackage{tabularx}
\usepackage{amsmath}
\usepackage{amssymb}
\usepackage{amsfonts}
\usepackage{graphicx}
\usepackage{xcolor}
\usepackage{microtype}
\usepackage{nicefrac}
\usepackage{enumitem}
\usepackage{caption}
\usepackage{subcaption}
\usepackage{adjustbox}
\usepackage{pgfplots}
\pgfplotsset{compat=1.18}
\titlespacing*{\paragraph}{0pt}{6pt}{1em}

\newcommand{\R}{\mathbb{R}}
\title{When Model Complexity Matters in Cross-Subject Naturalistic fMRI Prediction}

\author{%
\parbox[t]{\dimexpr\textwidth-2\tabcolsep\relax}{%
\centering
\normalfont\normalsize
\textbf{%
Junhao Zhao\textsuperscript{1},
Jacob Kang\textsuperscript{2},
Colin Kurniawan\textsuperscript{3},
David Simberg\textsuperscript{3},
Nan Xu\textsuperscript{1,2,4,5}\thanks{Corresponding author: Nan Xu (\texttt{nanxu@umd.edu}). Emails (in author order): \texttt{\{jzhao121, jkang115, ckurniaw, dsimberg, nanxu\}@umd.edu}.}%
}\\[6pt]
\normalsize
\mbox{\textsuperscript{1}Department of Electrical and Computer Engineering}\quad
\mbox{\textsuperscript{2}Fischell Department of Bioengineering}\\[2pt]
\mbox{\textsuperscript{3}Department of Computer Science}\quad
\mbox{\textsuperscript{4}Brain and Behavior Institute}\\[2pt]
\mbox{\textsuperscript{5}Artificial Intelligence Interdisciplinary Institute at Maryland}\\[2pt]
University of Maryland, College Park, MD, USA\\[4pt]
}%
}

\date{}

\begin{document}

\maketitle
\pagestyle{plain}

\begin{abstract}
Increasingly complex neural models predict blood-oxygen-level-dependent (BOLD) responses to naturalistic stimuli, but the predictive value of this added complexity remains unclear. We ask when pretrained stimulus representations help and how much downstream complexity is needed, with and without recent measured BOLD history. We compare linear diagnostic references and nonlinear predictors across HCP movie fMRI (154 subjects), BOLD Moments (10 subjects), and Algonauts 2025 (4 subjects). Under a shared offline protocol, models are fitted on training subjects and evaluated on held-out subjects viewing largely familiar stimuli, without refitting. Pretrained visual features generally improve prediction over raw pixels; for linear autoregression, this advantage becomes more pronounced over longer rollouts (recursive predictions without additional measured BOLD). Measured BOLD history provides substantial gains that generally diminish {as rollouts lengthen}. Nonlinear stimulus-only predictors can improve on linear readouts, whereas linear autoregression provides a strong history-conditioned baseline. Controls examine context length, regularization, fitting procedures, and model capacity. Regional response consistency across subjects is associated with held-out prediction accuracy, while less consistent regions tend to gain more from measured history. In paired simulations with fixed linear neural dynamics, slower hemodynamic responses generally sustain high temporal prediction correlations over longer {rollouts}, illustrating how the observation process can contribute to predictive persistence. Together, these findings support evaluating stimulus representations and downstream complexity separately, using history-conditioned linear prediction as a diagnostic reference under matched inputs and measurement availability.
\end{abstract}
\section{Introduction}
\label{sec:introduction}
Naturalistic fMRI prediction frameworks are becoming increasingly complex, but it remains unclear where that complexity actually pays off. Movies combine visual, auditory, and linguistic information over time, eliciting distributed and dynamic BOLD responses \citep{hasson2004intersubject}. Recent brain foundation models and related neural architectures bring richer stimulus representations and increasingly sophisticated predictors to this problem \citep{dascoli2025tribe,dascoli2026foundation,gokce2026mirage,yin2026brainvista,dong2026neuroworld}. We ask a simple question: \emph{where and when is this additional model complexity actually useful?}

Cross-subject prediction makes this question especially meaningful. People watching the same movie share stimulus-locked responses, yet their BOLD trajectories differ across individuals \citep{hasson2004intersubject,finn2020idiosynchrony}. Predicting an unseen subject therefore remains challenging \citep{wen2018transferring,dascoli2026foundation}. Transfer learning and target-subject adaptation reduce data requirements but still require subject-specific fitting \citep{wen2018transferring,mai2025synbrain}, which may be impractical when paired naturalistic stimulus--fMRI data are scarce. We ask whether recent measured BOLD can inform predictions for held-out subjects without refitting. This is neuroscientifically plausible because spontaneous BOLD activity carries individual-specific structure and contributes to variability in task-evoked responses \citep{tavor2016task,fox2006coherent}.

\begin{figure}[t]
\centering
\includegraphics[width=380pt]{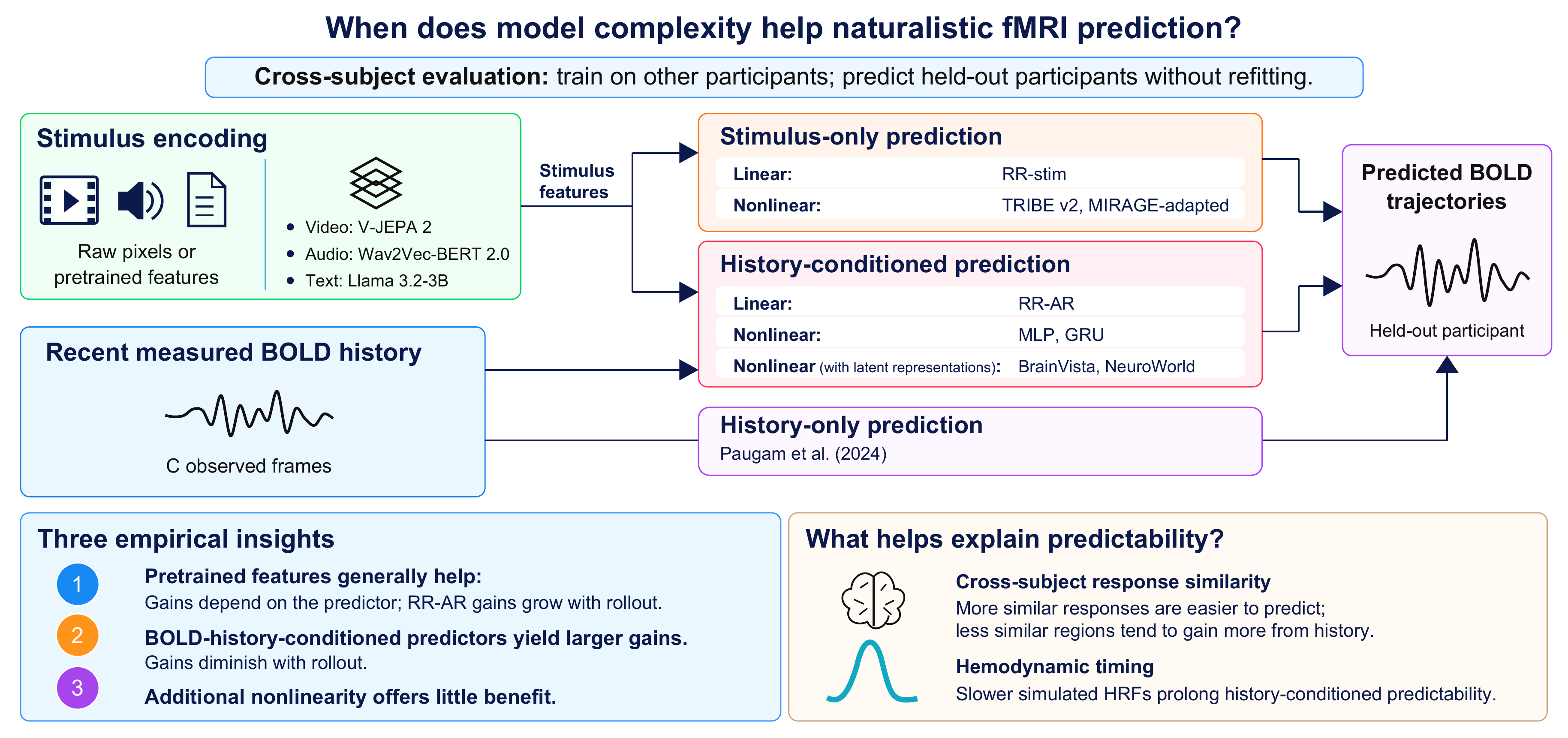}
\caption{Overview of cross-subject naturalistic fMRI prediction and main findings.}
\label{fig:overview}
\end{figure}

Naturalistic fMRI predictors use stimulus inputs, measured BOLD history, or both (Figure~\ref{fig:overview}). \emph{Stimulus-only} models, such as TRIBE v2~\citep{dascoli2026foundation} and MIRAGE~\citep{gokce2026mirage}, predict BOLD from stimulus representations without observing the target subject's recent BOLD responses. \emph{History-only} models predict future BOLD solely from recent BOLD responses, without stimulus inputs. \emph{History-conditioned} models incorporate both stimulus information and recent measured BOLD time series as model inputs, as in BrainVista~\citep{yin2026brainvista} and NeuroWorld~\citep{dong2026neuroworld}. Although not designed specifically for no-refit cross-subject prediction, these formulations naturally admit recent BOLD as subject-specific input. Across these predictors, we distinguish two components of model complexity: \emph{stimulus encoding}, for models with stimulus inputs, and \emph{downstream prediction}. Stimulus-encoding complexity is illustrated by TRIBE v2's pretrained video, audio, and language encoders~\citep{dascoli2026foundation} and MIRAGE's multimodal backbone with learned feature aggregation~\citep{gokce2026mirage}. Downstream complexity includes recurrent architectures such as GRU and learned latent dynamics such as those in NeuroWorld~\citep{dong2026neuroworld}.

These two sources of model complexity serve different roles. Richer stimulus representations may better capture the sensory, semantic, and temporal structure of naturalistic experience~\citep{huth2012continuous,deheer2017hierarchical}. Downstream predictors model BOLD, a noisy macroscopic signal shaped by neurovascular coupling and hemodynamic filtering~\citep{logothetis2001neurophysiological,nozari2024macroscopic}. The complexity of neural computation therefore does not directly determine the model complexity needed to predict BOLD trajectories~\citep{nozari2024macroscopic}. For example, \citet{paugam2024benchmark} compared linear and nonlinear history-only models during movie viewing and found linear models competitive with nonlinear alternatives.

{To determine where complexity pays off, we compare raw pixels with pretrained visual features, linear with nonlinear downstream predictors, and prediction from stimulus alone, measured BOLD history alone, or both. Our diagnostic references are stimulus-only ridge regression (RR-stim) and history-conditioned ridge autoregression (RR-AR). Nonlinear comparisons include stimulus-only TRIBE v2 and MIRAGE, and history-conditioned MLP, GRU, BrainVista, and NeuroWorld; the MLP and GRU match RR-AR's parameter count. We evaluate all models on HCP movie fMRI, BOLD Moments, and Algonauts 2025 under a shared cross-subject protocol without refitting on held-out subjects. We then relate regional prediction accuracy and history gains to cross-subject response similarity. We further construct controlled audiovisual simulations that transform neural activity into BOLD through an explicit biophysical hemodynamic model. These provide a mechanistic test bed for asking how the fMRI measurement itself shapes the persistence of history-conditioned predictability.}

Across these controlled comparisons, \textbf{three key insights emerge}:
\begin{itemize}[leftmargin=*,topsep=0pt,itemsep=2pt,parsep=0pt,partopsep=0pt]
    \item \textbf{Pretrained stimulus representations generally improve prediction.} Gains over raw pixels are clearest for stimulus-only predictors and RR-AR, whose gains increase with rollout length.\footnote{A \emph{rollout} consists of recursive predictions without additional measured BOLD. A \emph{refresh} updates the prediction history with the latest measured BOLD frames; the refresh interval $H$ is the number of prediction steps between these updates.}
    \item \textbf{BOLD history yields large prediction gains at shorter refresh intervals.} History-only controls show that recent BOLD alone supports strong short-horizon prediction, whereas stimulus input generally adds more value as refresh intervals lengthen.
    \item \textbf{The value of downstream nonlinearity depends on history availability.} Nonlinear stimulus-only predictors can outperform linear readouts, whereas RR-AR generally leads the tested history-conditioned comparisons.
\end{itemize}
Regional analyses further relate predictability to cross-subject response similarity, while controlled simulations show that slower hemodynamic time scales can prolong the predictive value of recent history. Together, these results clarify when model complexity improves cross-subject naturalistic fMRI prediction and illustrate how the hemodynamic observation process can prolong history-conditioned predictability.
\section{Background and Related Work}
\label{sec:background}
\textbf{Stimulus encoding.}
Visual encoding has progressed from hand-designed features to deep visual representations~\citep{kay2008identifying,wen2018neural}. TRIBE and TRIBE v2 combine pretrained video, audio, and language encoders, whereas MIRAGE uses a natively multimodal backbone with learned feature aggregation~\citep{dascoli2025tribe,dascoli2026foundation,gokce2026mirage}. Following TRIBE v2, we use V-JEPA~2, Wav2Vec-BERT~2.0, and Llama~3.2-3B for video, audio, and text, respectively~\citep{assran2025vjepa2,seamless2023communication,meta2024llama32}. We additionally evaluate two popular visual encoders, DINOv2~\citep{oquab2023dinov2} and CLIP (using an OpenCLIP checkpoint)~\citep{cherti2023reproducible}, in visual-only comparisons on HCP movie fMRI. These comparisons test whether pretrained visual features improve prediction over raw pixels across downstream predictors; a comprehensive study of encoder architectures and pretraining strategies remains beyond our scope.

\textbf{Downstream prediction.}
{Downstream predictors differ in their inputs and mappings to BOLD. Stimulus-only predictors map stimulus features to responses using linear readouts or nonlinear temporal models, while history-conditioned predictors additionally use recent measured BOLD. BrainVista tokenizes BOLD for autoregressive prediction, whereas NeuroWorld learns stimulus-conditioned latent dynamics and trains a decoder on latent rollouts } \citep{yin2026brainvista,dong2026neuroworld}. These history-conditioned architectures complement stimulus-only temporal predictors such as TRIBE v2 and MIRAGE (Appendix Table~\ref{tab:related_model_classes}).

\textbf{Linear prediction of BOLD dynamics.}
Linear models have performed strongly in resting-state prediction
\citep{nozari2024macroscopic} and movie-fMRI prediction with history input alone
\citep{paugam2024benchmark}. \citet{nozari2024macroscopic} examine averaging, observation noise, and finite samples as possible explanations, while \citet{paugam2024benchmark} find that linearized graph-based predictors outperform their nonlinear counterparts. {We build on these findings by testing whether downstream nonlinearity adds predictive value when pretrained stimulus representations and measured BOLD history are combined for cross-subject prediction.}

\section{Experimental Setup}%
\label{sec:experiments}
\subsection{Predictors and diagnostic references}
\label{sec:baselines}
We compare eight predictors (Figure~\ref{fig:overview}; Table~\ref{tab:related_model_classes}): 3 stimulus-only models use stimulus features alone, and 5 history-conditioned predictors also use the latest $C$ measured BOLD frames to predict the next $H$ steps before refreshing their history input with newly measured BOLD. The comparison includes diagnostic references and recent neural architectures adapted to a shared cross-subject protocol.
\paragraph{Linear and nonlinear diagnostic references.}
We construct two linear references. Following the linear readout approach used in stimulus encoding~\citep{wen2018neural}, stimulus-only ridge regression (RR-stim) maps stimulus features directly to BOLD. Stimulus-conditioned ridge autoregression (RR-AR) additionally incorporates recent BOLD history. Let $\mathbf y_t\in\mathbb R^R$ denote the parcel-level BOLD vector and $\mathbf x_t\in\mathbb R^D$ the aligned stimulus representation. RR-AR predicts 
\begin{equation}
\hat{\mathbf y}_t=\mathbf b+\sum_{\ell=1}^{C}A_\ell\tilde{\mathbf y}_{t-\ell}+B\mathbf x_t,\qquad A_\ell\in\mathbb R^{R\times R},\quad B\in\mathbb R^{R\times D}.
\label{eq:rr}
\end{equation}
Here $C$ is the history length, $\tilde{\mathbf y}$ contains measured BOLD within the supplied context and predicted BOLD thereafter, and each $A_\ell$ allows cross-parcel prediction. RR-stim omits the history terms and is fitted separately. Both models minimize one-step squared error with an $\ell_2$ penalty and an unpenalized intercept. Penalties are selected by multi-step validation BOLD MSE; fitting details are in Appendix~\ref{app:bold_baseline_details}.

To test additional downstream nonlinearity, we replace the linear mapping with an MLP or GRU receiving the same stimulus features and BOLD-history window as RR-AR. Both directly predict parcel-level BOLD trajectories and feed their predictions back during rollout. Their parameter counts match RR-AR within 1\%, and they are trained on multi-step BOLD prediction error.

\paragraph{Recent neural models and benchmark adaptations.}
We evaluate stimulus-only TRIBE v2 and MIRAGE, and history-conditioned BrainVista and NeuroWorld under the shared cross-subject protocol. TRIBE v2 and MIRAGE use released downstream code; BrainVista and NeuroWorld are implemented based on their published descriptions, retaining their learned BOLD representations and staged training procedures. All downstream neural models are initialized from scratch and trained on the common frozen stimulus features. For MIRAGE, we replace its native Qwen-Omni representations and subject-specific heads with the common features and a shared parcel readout. Model names throughout refer to these benchmark implementations. Model-specific pretraining objectives are retained where applicable, while final prediction checkpoints are selected using the same complete-coverage multi-step validation BOLD MSE used to select ridge penalties. Stimulus features are extracted offline, and TRIBE v2 and MIRAGE use bidirectional attention within the supplied stimulus windows. Feature extraction and temporal alignment are detailed in Appendix~\ref{app:stimulus_delay}, while architectures, adaptations, and training schedules are described in Appendix~\ref{app:implementation}.

\subsection{Datasets and stimulus representations}
We evaluate cross-subject prediction on three naturalistic fMRI datasets to test whether the predictive value of model complexity generalizes across subject sample sizes, stimulus types, and stimulus durations: HCP movie fMRI~\citep{hcp7tprotocol}, BOLD Moments~\citep{lahner2024boldmoments}, and Algonauts 2025~\citep{gifford2025algonauts}. These provide complementary settings, spanning four movie runs in a large subject cohort, short visual events, and extensive television/movie recordings from a small cohort. For HCP and BOLD Moments, BOLD time series are extracted using the Schaefer-Yeo 1,000-parcel atlas~\citep{schaefer2018localization}; Algonauts retains its native 1,000-parcel targets.  Table~\ref{tab:datasets0} summarizes the dataset designs and cross-subject splits. {
Evaluated BOLD duration is the time span used to compute prediction accuracy per clip or run. OFF and ON denote durations without and with video presentation, respectively. BOLD Moments includes 1-s inter-trial intervals, a 4-s initial fixation, and 4-s null trials. Algonauts presents continuous video within runs; between-run break durations are unreported. Appendix Table~\ref{tab:datasets} provides additional acquisition and preprocessing details.
\begin{table}[htbp]
\centering
\small
\caption{Dataset designs and cross-subject splits
(HCP: train/validation/test; LOSO: leave-one-subject-out).
Full details are in Appendix~\ref{app:data}.}
\label{tab:datasets0}
\setlength{\tabcolsep}{3pt}
\renewcommand{\arraystretch}{1.05}

\begin{adjustbox}{max width=\linewidth}
\begin{tabular}{@{}lccccccc@{}}
\toprule
Dataset
& Subjects
& \shortstack{Subject split}
& \shortstack{TR (s)}
& \shortstack{Unique movie\\duration}
& \shortstack{Train--test\\stimulus overlap}
& \shortstack{Evaluated BOLD\\duration (min)}
& {stimuli OFF / ON} \\
\midrule
HCP four movies
& 154
& 107/15/32
& 1.00
& 49.4 min
& 100\%
& 1.1--4.3 / clip
& {20 s / 1.1--4.3 min} \\

BOLD Moments
& 10
& \shortstack{5 two-subject folds}
& 1.75
& 5.1 min
& 100\%
& 7.5 / run
& {1 s / 3 s} \\

Algonauts 2025
& 4
& 4 LOSO folds
& 1.49
& 63.1 h
& 89.7\%
& 9.0--14.4 / clip
& {n.a / 10--12 min} \\
\bottomrule
\end{tabular}
\end{adjustbox}
\par\smallskip

\end{table}

Following TRIBE's feature-extraction pipeline~\citep{dascoli2025tribe}, all benchmark methods share cached features from frozen V-JEPA2~\citep{assran2025vjepa2}, Wav2Vec-BERT 2.0~\citep{seamless2023communication}, and Llama 3.2-3B~\citep{meta2024llama32} encoders. In our benchmark, features are aligned to the BOLD grid with a fixed three-step lag and standardized using training data. BOLD Moments subjects viewed silent videos: its audio and ASR-text features describe the source videos, not additional presented sensory inputs. Preprocessing, feature dimensions, missing-modality handling, and alignment are specified in Appendix~\ref{app:data}.

Downstream predictors are trained separately for each dataset using the shared stimulus representations, with parameters learned from training subjects and no refitting on held-out subjects. Stimulus content can recur across training and test subjects; the benchmark evaluates generalization to new subjects viewing familiar stimuli.

\subsection{Controlled audiovisual simulation}
\label{sec:simulation_setup}
To examine how the biophysical transformation from neural activity to BOLD shapes predictability, we adapt the DCM--Balloon framework of \citet{smith2011network} to a simplified 11-ROI audiovisual circuit (Fig.~\ref{fig:flat}(a)). Synthetic visual and auditory inputs enter V1 and A1, respectively, while stochastic background inputs drive all ROIs throughout fixation and stimulation. A fixed network generates linear neural dynamics, which a nonlinear Balloon--Windkessel model transforms into BOLD~\citep{friston2000nonlinear}. Each five-minute recording begins with 20 seconds of fixation, followed by sustained stimulation or intermittent stimulation at 5 s on/25 s off or 20 s on/20 s off. For each subject and schedule, the same neural trajectory generates BOLD under HRFs with target times to peak of 2, 6, and 10 seconds  (Fig.~\ref{fig:flat}(b)). Within each fast, medium, or slow condition, all ROIs share a common nominal HRF. Neural dynamics are simulated with a 5-ms time step and passed through the Balloon--Windkessel model. The resulting BOLD signals are sampled at $\mathrm{TR}=1$~s for prediction. The matched no-HRF neural reference is downsampled to 1 TR. We simulate 50 subjects per schedule, use a fixed 35/5/10 train/validation/test split, and give predictors the synthetic stimulus directly, without pretrained encoders. Further settings are in Appendix~\ref{app:hrf_model_controls}.

\subsection{Evaluation protocol and metrics}
\label{sec:evaluation}

\paragraph{Prediction protocol.}
History-conditioned predictors start with $C=8$ measured BOLD frames and predict one frame at a time, feeding each prediction into the rolling history. We call this recursive prediction without additional measured BOLD a \emph{rollout}. After $H$ predictions, the history is refreshed with the latest $C$ measured BOLD frames. For example, with $H=4$, measured frames 1--8 initialize predictions for frames 9--12. Once frames 9--12 are measured, frames 5--12 form the history for predicting frames 13--16. Thus, $H$ controls the number of prediction steps between measured-BOLD updates, corresponding to $H$ TRs. We evaluate a range of refresh intervals and a no-refresh condition that uses measured BOLD only in the initial context. For each ROI within each clip or scan run, we correlate the predicted and measured BOLD time series over the scored samples. For HCP, preceding fixation supplies the initial history and all within-clip BOLD samples are scored; for BOLD Moments and Algonauts, scoring starts after the initial eight aligned BOLD frames (Appendix~\ref{app:data}). Each clip or run is predicted independently, and the same target samples are scored at every refresh interval. Stimulus-only predictions are independent of $H$. Further details of the rollout and refresh protocol, including the full refresh grid, boundary conventions, and a schematic illustration, are provided in Appendix~\ref{app:prediction_protocol} (Figure~\ref{fig:rollout_protocol}). We report downstream predictor parameters or fitted ridge coefficients, including intercepts but excluding frozen stimulus encoders (Table~\ref{tab:parameter_counts}). These counts describe model size rather than training or inference cost.

\paragraph{Evaluation metrics.}
To evaluate trajectory accuracy, our primary metric is the temporal Pearson correlation between predicted and measured BOLD for each parcel within each clip or scan run~\citep{gifford2025algonauts}. We further assess inter-parcel correlation structure using functional connectivity (FC) at $H=32$. For predicted and measured time series, we compute pairwise parcel correlations within each record and average them across records within each subject using Fisher-$z$ averaging. FC similarity is the Pearson correlation between the off-diagonal upper-triangular entries of the predicted and measured subject-level FC matrices. Seed-specific similarities are averaged within subject, and we report the mean and sample SD across held-out subjects. We also examine mean FC and intersubject FC variability, quantified as the SD of FC entries across subjects, with parcels grouped by the 7-Yeo networks \citep{ThomasYeo2011TheOO}. Results include all designated folds and neural seeds 0/1/2; ridge fits are deterministic. Error-bar definitions are specified in the figure captions, with aggregation details in Appendices~\ref{app:benchmark_evaluation} and~\ref{app:regional}.

\section{Results}
\label{sec:results}
{\color{black}
}

\subsection{Stimulus representation: pre-trained stimulus representations improve prediction}
\label{sec:representation_gain}

Stronger visual representations benefit most of the tested predictors. Compared with raw pixels, V-JEPA 2 features improve the performance of most of the eight predictors across four HCP movies. Figure~\ref{fig:matched_visual_inputs}a presents three examples, with RR-stim and TRIBE v2 showing the largest relative gains on $H=never$ (+35\% and +13\%, respectively). Results for the other models are shown in Figure~\ref{fig:matched_visual_all_app}: BrainVista changes little or declines slightly on HCP, while most comparisons on BOLD Moments and Algonauts 2025 show the same favorable trend. Separate comparisons using DINOv2 and CLIP encoders reveal a similar overall pattern (Appendix~\ref{app:general_stimenc}). 

\begin{figure}[h!]
\centering
\includegraphics[width=\linewidth]{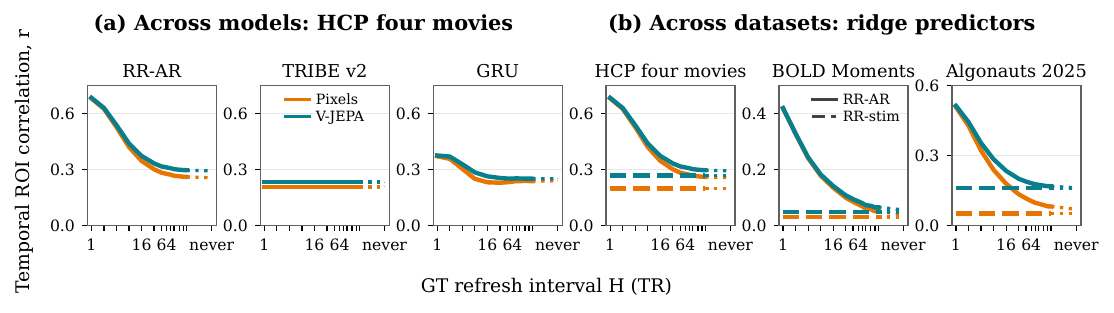}
\caption{Pixels versus V-JEPA visual representations across representative models on HCP four movies (a) and ridge predictors across three datasets (b). Results include all designated folds and neural seeds 0/1/2; complete comparisons are provided in Appendix~\ref{app:visual_representation_all}.}
\label{fig:matched_visual_inputs}
\label{fig:cross_model_pixels}
\label{fig:capacity_representation}
\label{fig:encoder}
\label{fig:ridge_stimulus_representation}
\end{figure}

The size of the V-JEPA gain varies across predictors and datasets. At the no-refresh endpoint, all eight predictors show their largest absolute gain over pixels on Algonauts; MIRAGE gains 0.084 in correlation there, versus 0.007 on HCP and 0.005 on BOLD Moments (Appendix Figure~\ref{fig:matched_visual_all_app}). For RR-AR, the advantage becomes more apparent as measured-BOLD refreshes become less frequent (Figure~\ref{fig:capacity_representation}b). Separate modality controls also evaluate video, audio, and text alone and in combination (Appendix~\ref{app:modality}). These results support the predictive value of the tested visual representations across datasets. Feature dimensions and preprocessing differ across representations, so these comparisons evaluate
representation pipelines without isolating the effect of pretraining under dimension-matched inputs (Appendices~\ref{app:representation_controls}
and~\ref{app:general_stimenc}).

\subsection{Predictor Performance}
\label{sec:history_benefit}
\label{sec:history_fading}
\label{sec:performance}
\label{sec:rr_benchmark}

\begin{figure}[!htbp]
    \centering
    \includegraphics[width=\linewidth]{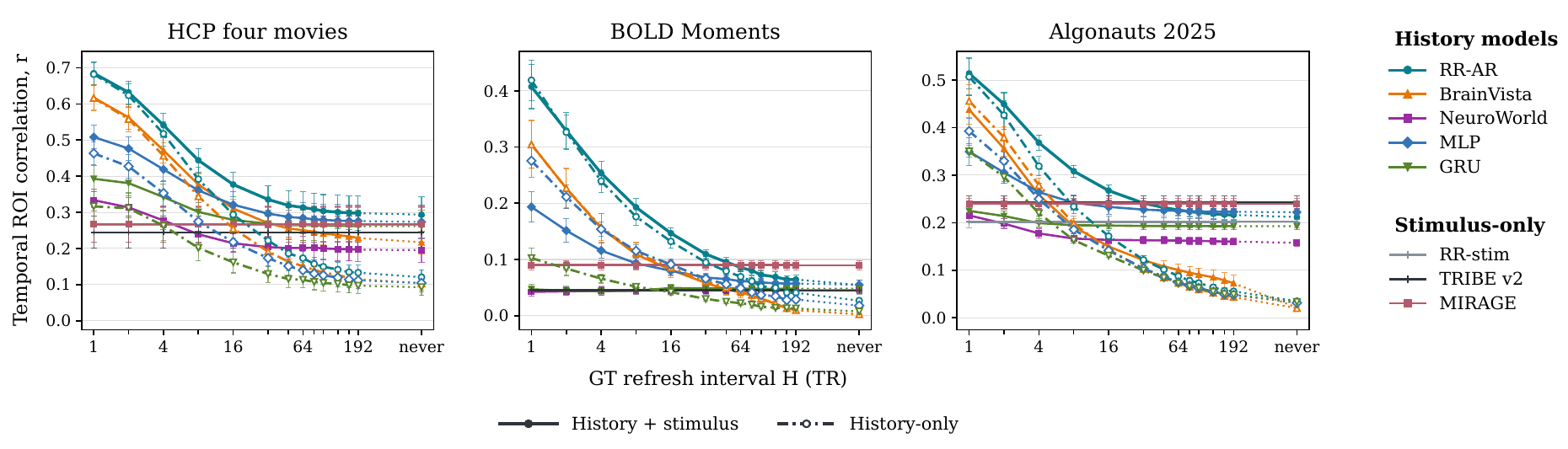}
    \caption{Cross-subject fMRI prediction across three datasets for eight predictors and history-only refits of RR-AR, BrainVista, MLP, and GRU. Error bars show SD across held-out subjects. Complete ablation results and the rationale for excluding NeuroWorld's history-only control are provided in Appendix~\ref{app:history_only_controls}.}
    \label{fig:baseline}
    \label{fig:controlled_absolute}
\end{figure}

\paragraph{Recent BOLD supports short-interval prediction, while stimulus input adds value at longer intervals.}
\label{sec:history_only_main}
At short refresh intervals ($H\leq4$), RR-AR, BrainVista, and MLP outperform all three stimulus-only references across all three datasets (Figure~\ref{fig:baseline}; Appendix Figure~\ref{fig:history_stimulus_main}). History-only RR-AR nearly matches the full model in this regime, indicating that recent BOLD accounts for much of its predictive accuracy. As refresh intervals lengthen, history-only RR-AR accuracy declines, while stimulus-only predictions remain unchanged. At longer refresh intervals, adding stimulus input yields larger gains over history-only RR-AR on HCP and Algonauts than on BOLD Moments, potentially reflecting their longer, more continuous movie stimuli compared with the brief 3-s videos separated by 1-s fixation intervals in BOLD Moments.

\paragraph{Nonlinearity can improve stimulus-only prediction.} Among stimulus-only predictors, MIRAGE\ performs similarly to RR-stim on HCP, differing by about 1\%, but clearly outperforms it on BOLD Moments and Algonauts. These results show that nonlinear stimulus-to-BOLD mappings can improve prediction when models use stimulus features alone, although the gains vary across datasets.

\paragraph{Linear prediction leads most history-conditioned comparisons.}
\label{sec:complexity}
\label{sec:rr_rank_diagnostic}
\label{sec:diagnostic_factors}
\label{sec:rr_robustness_diagnostic}
With comparable stimulus features and BOLD-history inputs, RR-AR generally outperforms the nonlinear alternatives, including the parameter-matched MLP and GRU controls. On one diagnostic split per dataset, RR-AR remains competitive across the tested context lengths and neural training intervals (Appendix Figures~\ref{fig:context_absolute} and~\ref{fig:training_k_absolute}). Absolute accuracy is sensitive to ridge regularization (Appendix~\ref{app:regularization_curves}). On the same diagnostic splits, mini-batch optimization of the one-step ridge objective closely reproduces the closed-form model's prediction accuracy (Appendix~\ref{sec:rr_fitting_diagnostic}). Reduced-rank variants, evaluated on all designated held-out subjects, retain much of RR-AR's performance with substantially fewer coefficients (Appendix~\ref{app:reduced_rank_controls}). Together, these controls support the strength of linear history-conditioned prediction across the tested fitting procedures and capacity choices, while retaining sensitivity to hyperparameters and evaluation conditions.

\subsection{Regional analysis: where does measured history help linear prediction?}
\label{sec:shared_results}
We next examine regional differences in prediction accuracy and whether gains from incorporating measured BOLD are greater where cross-subject response similarity is low. %

\paragraph{Measured BOLD history supports regional and network-level prediction.} At $H=32$, the displayed models predict visual and auditory cortex comparatively well, while RR-AR retains higher accuracy in dorsal frontoparietal and medial association cortex, broadly overlapping Dorsal Attention, Control, and Default networks (Figure~\ref{fig:regional-fc}, top). On HCP, BrainVista also predicts these association regions more accurately than MIRAGE and GRU. This pattern is consistent with the longer temporal integration windows of association cortex~\citep{hasson2008hierarchy}, suggesting that recent BOLD may supply accumulated stimulus context and ongoing response information that stimulus features alone miss. Across all three datasets, incorporating measured BOLD history improves FC similarity within the linear ridge framework, with RR-AR exceeding RR-stim by 0.14, 0.19, and 0.09 on HCP, BOLD Moments, and Algonauts, respectively (Appendix Figure~\ref{fig:fc_h32_all} and Table~\ref{tab:fc_h32}). RR-AR achieves the highest mean FC similarity on HCP and BOLD Moments and remains close to BrainVista and MLP on Algonauts. However, preserving FC structure does not guarantee accurate BOLD trajectories: on Algonauts at $H=32$, BrainVista achieves the highest mean FC similarity but the lowest mean temporal prediction accuracy (Figure~\ref{fig:baseline} and Appendix Table~\ref{tab:fc_h32}). On HCP, history-conditioned predictors produce appreciable intersubject FC variability, whereas stimulus-only predictors show little to no variability across subjects receiving the same stimulus inputs (Figure~\ref{fig:regional-fc}, bottom, and Appendix Figure~\ref{fig:fc_h32_all}). 

\begin{figure}[h!]
\centering

\noindent\makebox[\linewidth][l]{%
\begin{minipage}[c]{0.035\linewidth}
\centering
\rotatebox{90}{%
  \scriptsize\bfseries
  \shortstack{Parcel-level\\prediction accuracy}%
}
\end{minipage}%
\begin{minipage}[c]{0.845\linewidth}
\centering

\begin{subfigure}[t]{0.235\linewidth}
\centering
\caption{RR-AR}
\includegraphics[width=\linewidth]
{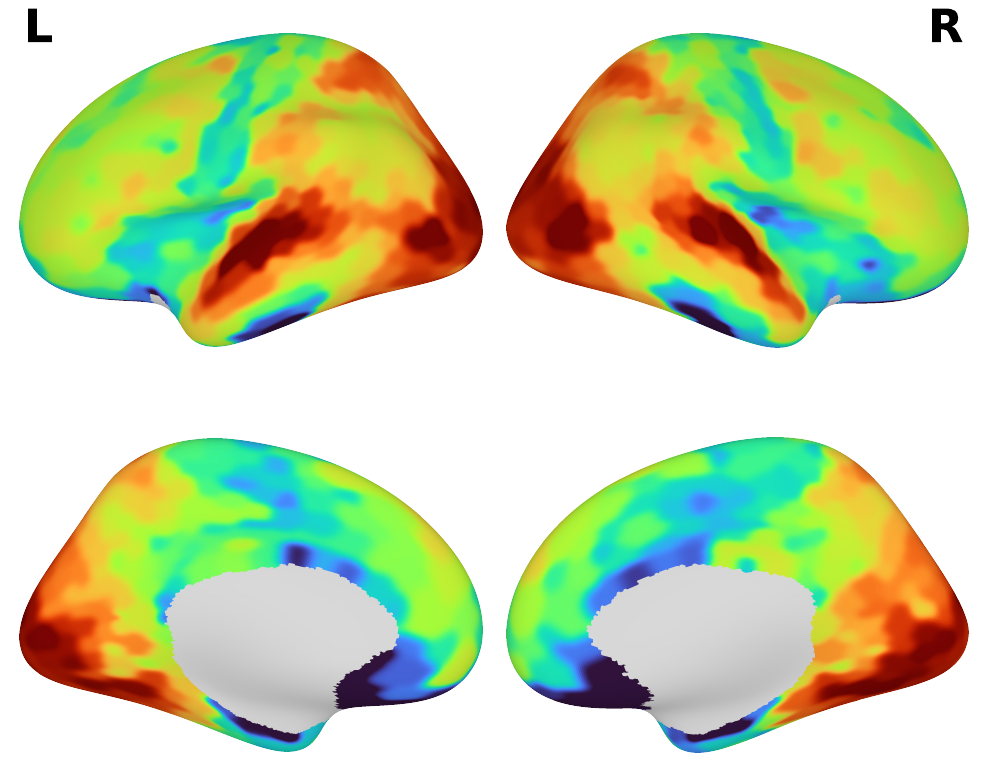}
\end{subfigure}\hfill
\begin{subfigure}[t]{0.235\linewidth}
\centering
\caption{BrainVista}
\includegraphics[width=\linewidth]
{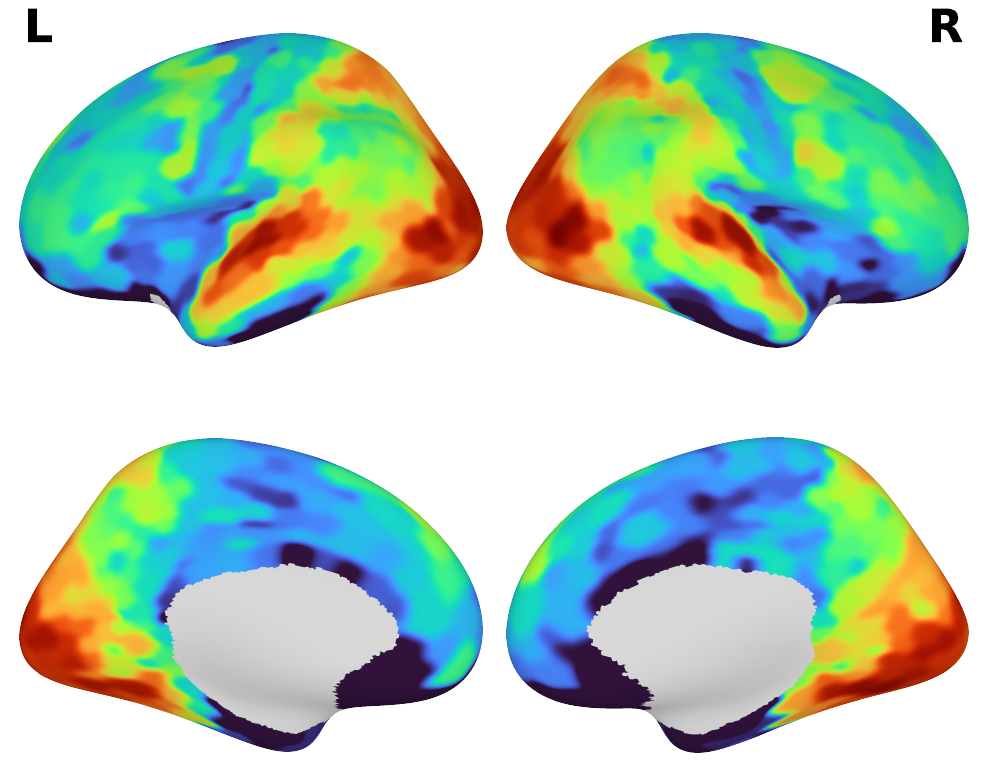}
\end{subfigure}\hfill
\begin{subfigure}[t]{0.235\linewidth}
\centering
\caption{MIRAGE}
\includegraphics[width=\linewidth]
{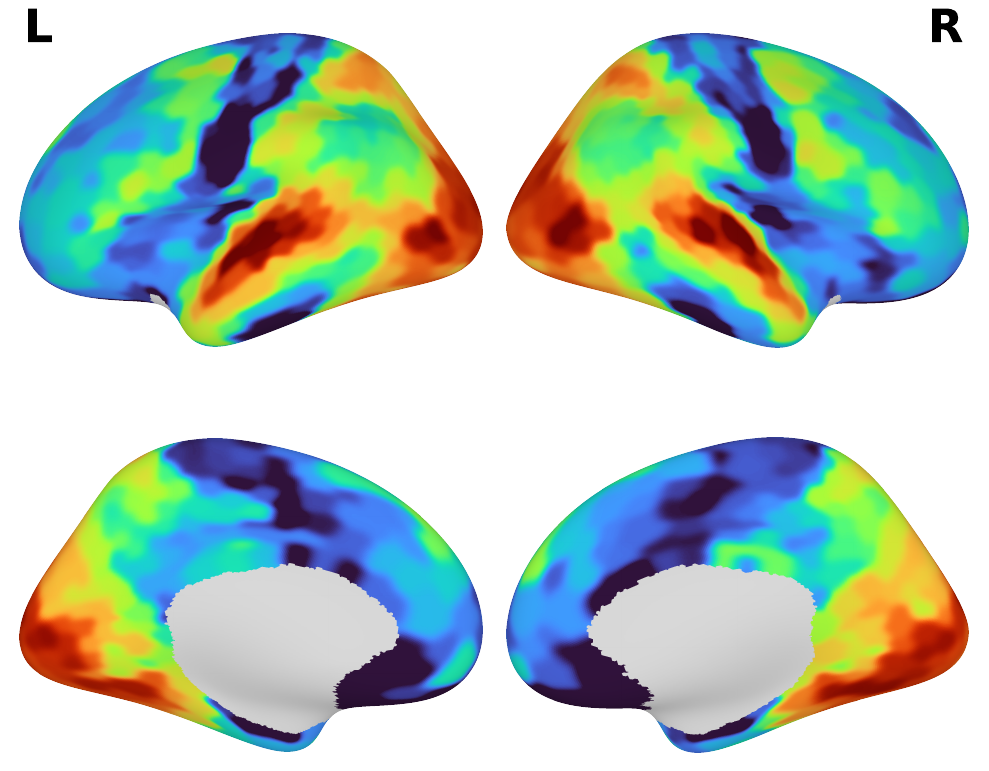}
\end{subfigure}\hfill
\begin{subfigure}[t]{0.235\linewidth}
\centering
\caption{GRU}
\includegraphics[width=\linewidth]
{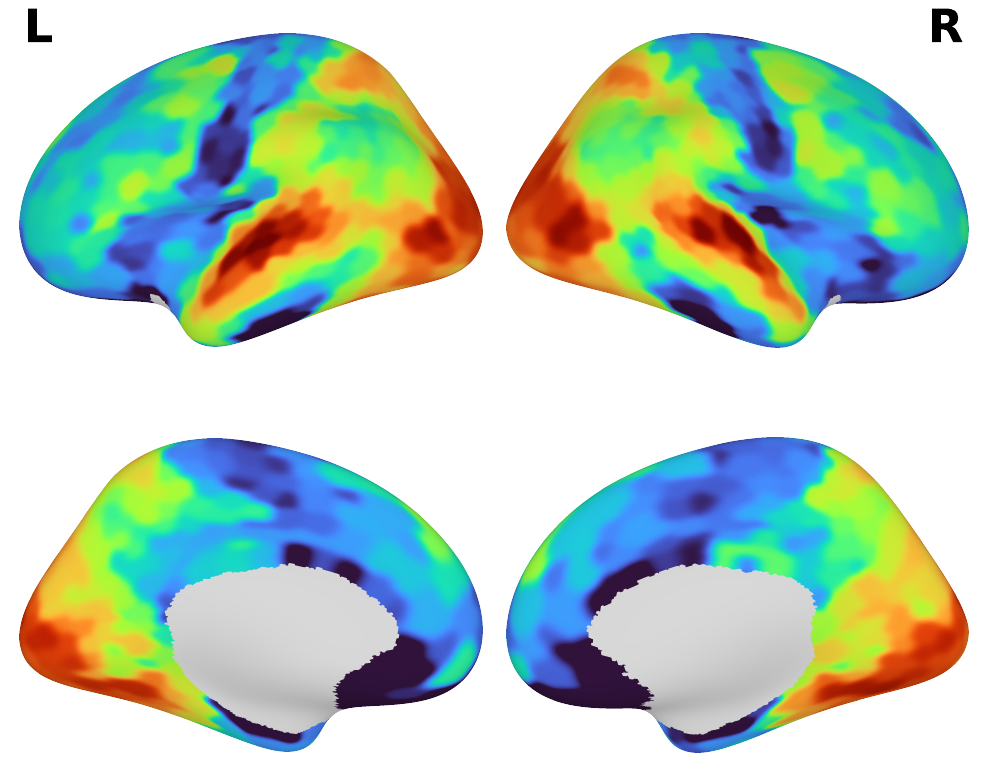}
\end{subfigure}

\end{minipage}\hfill
\begin{minipage}[c]{0.10\linewidth}
\centering
\includegraphics[
  height=0.125\textheight,
  keepaspectratio
]{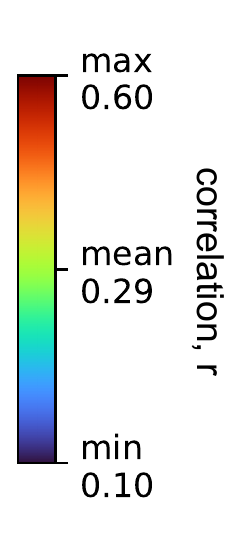}
\end{minipage}%
}

\par\smallskip

\noindent\makebox[\linewidth][l]{%
\begin{minipage}[c]{0.035\linewidth}
\centering
\rotatebox{90}{%
  \scriptsize\bfseries Average FC%
}
\end{minipage}%
\begin{minipage}[c]{0.845\linewidth}
\centering

\begin{minipage}[b]{0.235\linewidth}
\centering
\makebox[\linewidth][c]{%
  \tiny $r=0.730\mathbin{\pm}0.042$%
}
\par\nointerlineskip
\makebox[\linewidth][c]{%
  \includegraphics[width=1.04\linewidth]
  {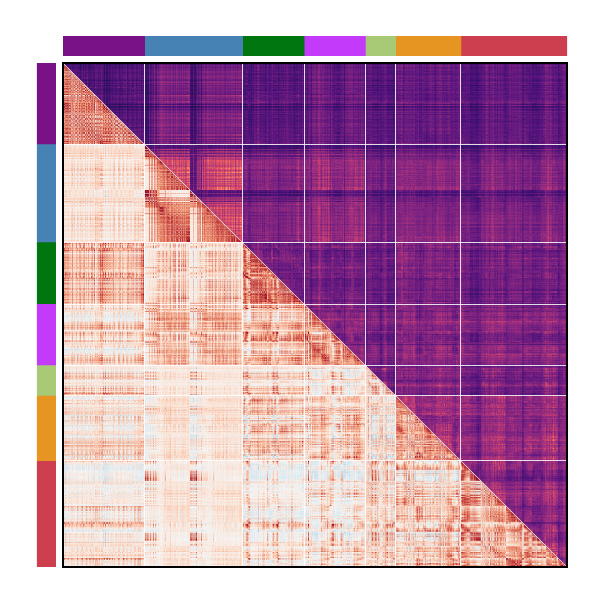}%
}
\end{minipage}\hfill
\begin{minipage}[b]{0.235\linewidth}
\centering
\makebox[\linewidth][c]{%
  \tiny $r=0.717\mathbin{\pm}0.033$%
}
\par\nointerlineskip
\makebox[\linewidth][c]{%
  \includegraphics[width=1.04\linewidth]
  {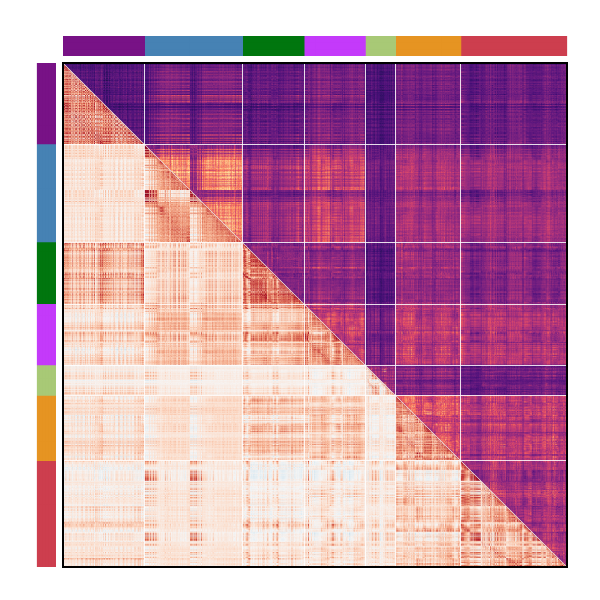}%
}
\end{minipage}\hfill
\begin{minipage}[b]{0.235\linewidth}
\centering
\makebox[\linewidth][c]{%
  \tiny $r=0.579\mathbin{\pm}0.052$%
}
\par\nointerlineskip
\makebox[\linewidth][c]{%
  \includegraphics[width=1.04\linewidth]
  {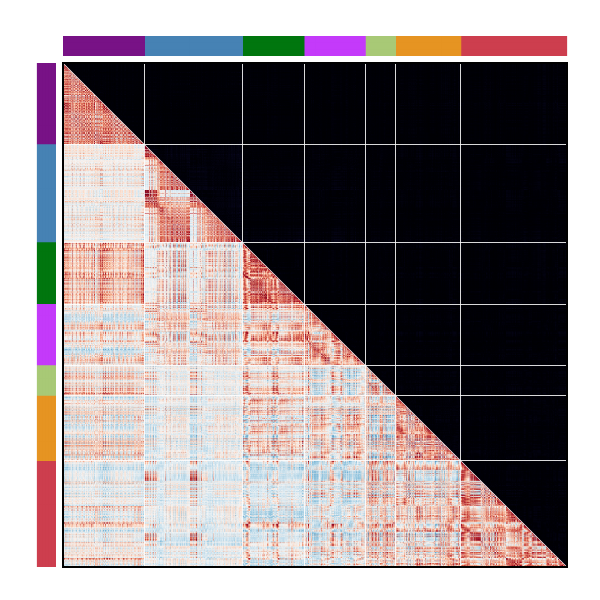}%
}
\end{minipage}\hfill
\begin{minipage}[b]{0.235\linewidth}
\centering
\makebox[\linewidth][c]{%
  \tiny $r=0.625\mathbin{\pm}0.046$%
}
\par\nointerlineskip
\makebox[\linewidth][c]{%
  \includegraphics[width=1.04\linewidth]
  {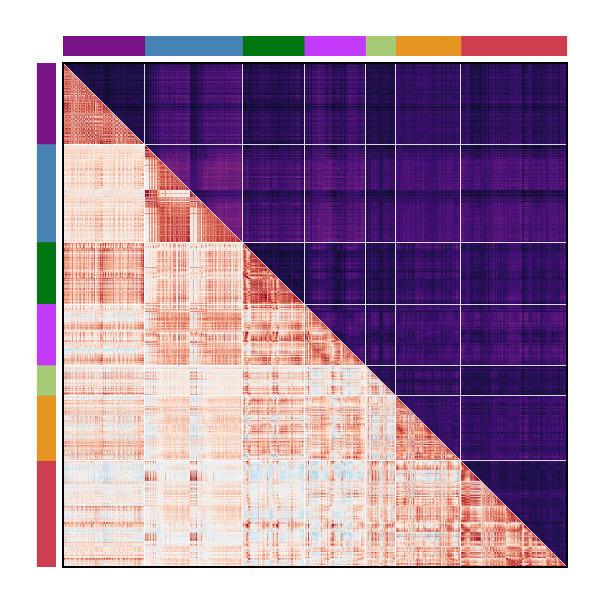}%
}
\end{minipage}

\end{minipage}\hfill
\begin{minipage}[c]{0.10\linewidth}
\centering
\includegraphics[
  height=0.125\textheight,
  keepaspectratio
]{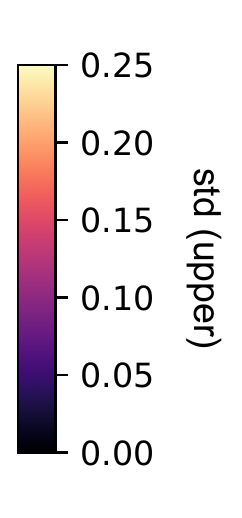}
\end{minipage}%
}

\par\nointerlineskip

\noindent
\begin{minipage}[t]{0.035\linewidth}
\mbox{}
\end{minipage}%
\begin{minipage}[t]{0.845\linewidth}

\begin{minipage}[t]{0.735\linewidth}
\raggedright
\includegraphics[width=\linewidth]
{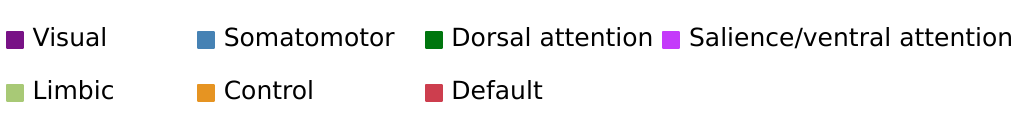}
\end{minipage}\hfill
\begin{minipage}[t]{0.235\linewidth}
\centering
\includegraphics[width=\linewidth]
{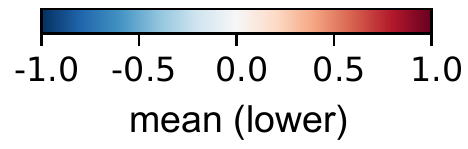}
\end{minipage}

\end{minipage}\hfill
\begin{minipage}[t]{0.10\linewidth}
\mbox{}
\end{minipage}
\caption{Regional prediction and functional connectivity (FC) on HCP at $H=32$. Top: parcelwise temporal prediction correlations. Bottom: mean predicted FC (lower triangles) and intersubject SD of Fisher-$z$ FC (upper triangles). Annotations report predicted--measured FC similarity (mean $\pm$ SD across held-out subjects). Full three-dataset results: Appendix~\ref{app:regional_fc}.\label{fig:regional-fc}}
\end{figure}

\paragraph{Measured BOLD history is particularly valuable in regions where responses differ more across individuals.}
\label{sec:rr_absolute_regional}
We ask whether regions that respond similarly across individuals are easier to predict, and whether regions with less similar responses benefit more from incorporating recent BOLD. We quantify response consistency across training subjects using temporal intersubject correlation (ISC) for HCP and Algonauts and event-aligned response-pattern correlation for BOLD Moments; higher values indicate more similar responses to the same stimulus (Appendix~\ref{app:regional_isc}). For each parcel, we compare this consistency with RR-AR's prediction accuracy (Figure~\ref{fig:regional_isc_h32}, left) and with the gain from incorporating BOLD history, defined as the difference in prediction correlation between RR-AR and RR-stim (Figure~\ref{fig:regional_isc_h32}, right). Across all three datasets, regions with more consistent responses are generally predicted more accurately. However, Control and Default networks are less accurately predicted by RR-stim than visual cortex, yet show larger gains from incorporating BOLD history. On HCP, these gains are 0.0445 in Control and 0.0417 in Default, nearly double the 0.0226 gain in the visual network. The same qualitative pattern persists at $H=96$: across parcels, response consistency is positively associated with prediction accuracy for both RR-AR and RR-stim and negatively associated with the gain from measured BOLD history in all three datasets (Appendix Table~\ref{tab:regional_isc_association}). This complementary profile is consistent with greater interindividual functional-connectivity variability in association cortex~\citep{mueller2013variability} and sensitivity to narrative context~\citep{yeshurun2017same}, suggesting that recent BOLD is particularly valuable where shared stimulus features provide a less complete account of the measured response. Because the consistency--accuracy relationship also holds for RR-stim, it characterizes general regional predictability rather than uniquely explaining the linear--neural performance gap (Appendix~\ref{app:regional_isc}).

\begin{figure}[!htbp]
\centering
\includegraphics[width=\linewidth]{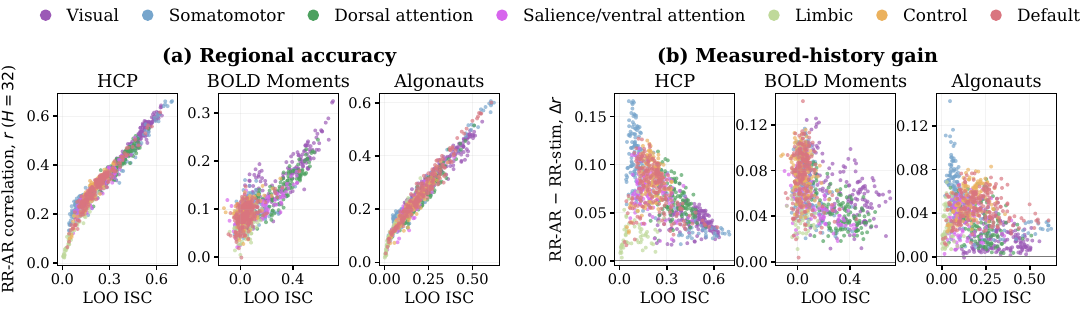}
\caption{Regional prediction accuracy and BOLD-history gains ($H=32$) versus training-subject response consistency. Each point represents a cortical parcel, with prediction performance evaluated on held-out subjects and response consistency (x-axis) computed across training subjects. (a) Absolute prediction accuracy for RR-AR. (b) Prediction gain of RR-AR over RR-stim. }
\label{fig:regional_isc_h32}
\end{figure}
\subsection{Hemodynamic response timing and history-conditioned predictability}
\label{sec:hrf_results}

Our empirical comparisons show that linear autoregression remains competitive with nonlinear predictors under matched stimulus and BOLD-history inputs (Figure~\ref{fig:baseline}). This finding is consistent with the strong linear predictability reported by \citet{nozari2024macroscopic}, whose analyses show that temporal averaging can mask underlying nonlinear dynamics. Hemodynamic filtering attenuates rapid neural fluctuations in the measured BOLD signal, effectively acting as a temporal low-pass filter~\citep{friston2000nonlinear,nozari2024macroscopic}. We therefore ask whether this filtering improves linear prediction and how its effect depends on the measured-BOLD refresh interval. To address this question, we conduct controlled simulations adapted from the DCM--Balloon framework of \citet{smith2011network}, varying HRF timing while holding simulated neural trajectories and stimulus inputs fixed. The simulated neural state before hemodynamic transformation provides a matched no-HRF reference (Section~\ref{sec:simulation_setup}).

\begin{figure}[!h]
\centering
\includegraphics[width=400pt]{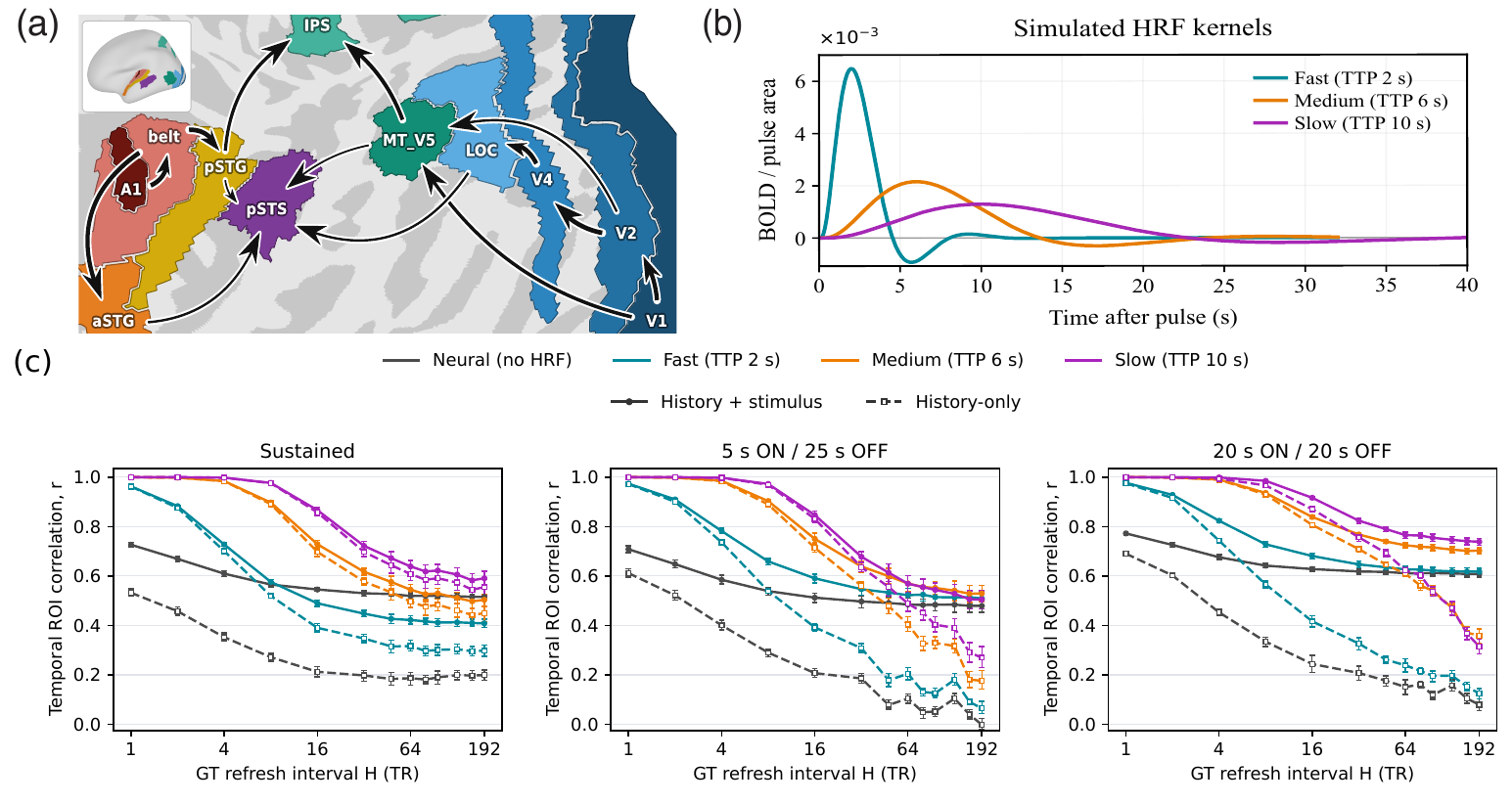}
\caption{Effect of hemodynamic response timing on RR-AR prediction. (a) Audiovisual circuit. (b) Fast, medium, and slow Balloon--Windkessel HRFs. (c) Prediction across refresh intervals and stimulus schedules for the no-HRF neural reference and HRF conditions. Solid lines show history-plus-stimulus predictions; dashed lines show history-only refits. Conditions share neural trajectories; $\mathrm{TR}=1$~s. Error bars: paired-subject bootstrap 95\% CIs. Details: Appendix~\ref{app:hrf_model_controls}.}
\label{fig:flat}
\label{fig:hrf_rr_absolute}
\end{figure}

Medium and slow HRF conditions produce a short-interval plateau with prediction correlations near one that is absent from the matched no-HRF neural reference (Figure~\ref{fig:hrf_rr_absolute}). Among the HRF conditions, slower HRFs generally sustain higher temporal prediction correlations over longer refresh intervals. These comparisons show that hemodynamic filtering can support sustained accurate linear prediction while the underlying neural trajectories remain unchanged.

RR-AR receives both stimulus inputs and BOLD history, so its strong prediction performance could be supported by either source of information. To determine whether BOLD history alone can support this performance, we refit a history-only RR-AR model with stimulus inputs zeroed. For the medium and slow HRF conditions, history-only prediction closely matches the full model at short refresh intervals and retains the BOLD prediction plateau, showing that recent BOLD alone supports much of this accuracy. Faster HRFs shorten the interval over which history alone sustains high accuracy. At longer intervals, history-only accuracy generally declines and the full model's advantage becomes more apparent, demonstrating the additional value of explicit stimulus inputs.

{We assess robustness across sustained stimulation and two ON/OFF schedules. The broad pattern of longer-lasting predictability with slower HRFs extends across these conditions, although its magnitude and the ordering of HRF conditions vary at longer intervals. The long-interval benefit of stimulus input is more pronounced under the ON/OFF schedules, whereas the full and history-only predictions remain closer under sustained stimulation, particularly for medium and slow HRFs. Complete results for all eight models and four history-only controls are provided in Appendix~\ref{app:hrf_all_models} (Figure~\ref{fig:hrf_all_models_appendix}).}

Together, these results show that the timing of the hemodynamic response affects how long recent BOLD history supports accurate linear prediction. By holding the underlying neural trajectories fixed, the paired simulations identify hemodynamic filtering as a contributor to this predictability.

\section{Discussion and Conclusion}
\label{sec:hrf_discussion}

\paragraph{Recent BOLD history provides subject-specific context for prediction.}
A stimulus-only predictor infers responses from stimulus features, whereas a history-conditioned predictor also observes recent responses from the target subject. History-only RR-AR retains much of the full model's accuracy at short refresh intervals, showing that recent BOLD already contains substantial predictive information. These observations may summarize accumulated stimulus effects and ongoing fluctuations, reducing how much the predictor must infer from shared stimulus features. The growing benefit of stimulus input at longer intervals indicates that measured history and stimulus representations provide complementary predictive information. This helps interpret why nonlinear mappings can improve stimulus-only prediction while linear autoregression remains strong when measured history is available. The regional analyses further suggest that measured history is particularly useful where responses are less consistent across subjects.

\paragraph{Hemodynamics shape the predictive value of BOLD history.}
Our controlled audiovisual simulations show that the timing of the hemodynamic response affects how long recent BOLD supports accurate prediction, even when the underlying neural trajectories are held fixed. History-only RR-AR retains the short-interval prediction plateau under medium and slow HRFs, demonstrating that this benefit persists without stimulus inputs supplied to the predictor. Temporal structure introduced by hemodynamics can therefore contribute directly to the predictive value of measured history. The corresponding HRF effects in nonlinear history-only predictors indicate that this contribution extends across model classes (Appendix Figure~\ref{fig:hrf_all_models_appendix}). Its dependence on stimulus scheduling further suggests that the usefulness of recent BOLD is jointly shaped by the observation process and the temporal structure of external inputs.

\paragraph{Implications for predictive modeling.}
Pretrained visual representations generally improve prediction across both linear and nonlinear predictors. RR-AR generally leads the tested history-conditioned comparisons and provides a strong diagnostic reference: within this model, pretrained visual features outperform raw pixels, with their advantage generally growing as measured-BOLD refreshes become less frequent. These results support evaluating stimulus encoding and downstream prediction separately. Stimulus-only, history-only, and combined-input controls further distinguish the contributions of stimulus representations and measured history. Additional downstream complexity should then be assessed against strong linear references under matched inputs and refresh intervals. The relevant question is whether that complexity improves prediction given the available information and measurement conditions.

\paragraph{Limitations.}
Our conclusions concern the tested benchmark adaptations and training procedures with shared frozen stimulus features, rather than the complete original pretrained systems. The benchmark primarily evaluates new subjects viewing familiar stimuli and uses offline preprocessing of complete recordings; generalization to unseen stimuli and prediction with online preprocessing remain to be established. The small subject cohorts in BOLD Moments and Algonauts also limit population-level conclusions. Finally, the simulations isolate hemodynamic effects in a simplified system with fixed linear neural dynamics; their generality to nonlinear neural dynamics remains to be tested.

\paragraph{Conclusion.}
Across three naturalistic fMRI datasets, pretrained stimulus representations generally improve prediction, and recent BOLD alone supports strong short-interval prediction. Stimulus input generally adds more value as measured-history refreshes become less frequent. Nonlinear stimulus-only predictors can improve on linear readouts, while linear autoregression remains a strong history-conditioned baseline. The paired simulations further identify hemodynamic filtering as a contributor to the predictive value of BOLD history. These findings support evaluating model complexity in relation to both available inputs and measurement conditions.

\Needspace{9\baselineskip}
\section*{Acknowledgements}
We acknowledge the use of generative AI tools for assisting with figure preparation, manuscript editing and organization, and code verification and reproducibility documentation; all final derivations, code, references, and content were thoroughly verified and approved by the authors, who take full responsibility for the paper. 

This research relies on secondary analysis of open-access human neurophysiological data from the Human Connectome Project (HCP) 7T release, collected with informed consent and institutional approval under its standard data-use terms; no restricted identifying variables were used, no re-identification was attempted, and the methodology developed herein is strictly experimental and not intended for clinical decision-making. 

Full dataset preprocessing details, train/test splits, model adaptations, hyperparameters, random seeds, and evaluation protocols are detailed in the Appendices to ensure complete reproducibility.

\bibliography{references}

@article{kay2008identifying,
  title={Identifying natural images from human brain activity},
  author={Kay, Kendrick N and Naselaris, Thomas and Prenger, Ryan J and Gallant, Jack L},
  journal={Nature},
  volume={452},
  number={7185},
  pages={352--355},
  year={2008},
  publisher={Nature Publishing Group UK London}
}

@article{ThomasYeo2011TheOO,
  title={The organization of the human cerebral cortex estimated by intrinsic functional connectivity.},
  author={B. T. Thomas Yeo and Fenna M. Krienen and Jorge Sepulcre and Mert R. Sabuncu and Danial Lashkari and Marisa Hollinshead and Joshua L. Roffman and Jordan W. Smoller and Lilla Z{\"o}llei and J. Polimeni and Bruce Fischl and Hesheng Liu and Randy L. Buckner},
  journal={Journal of neurophysiology},
  year={2011},
  volume={106 3},
  pages={
          1125-65
        },
  url={https://api.semanticscholar.org/CorpusID:263439953}
}

@article{oquab2023dinov2,
  title   = {{DINOv2}: Learning Robust Visual Features without Supervision},
  author  = {Oquab, Maxime and others},
  journal = {arXiv preprint arXiv:2304.07193},
  year    = {2023}
}

@inproceedings{cherti2023reproducible,
  title     = {Reproducible Scaling Laws for Contrastive
               Language-Image Learning},
  author    = {Cherti, Mehdi and Beaumont, Romain and Wightman, Ross
               and Wortsman, Mitchell and Ilharco, Gabriel
               and Gordon, Cade and Schuhmann, Christoph
               and Schmidt, Ludwig and Jitsev, Jenia},
  booktitle = {Proceedings of the IEEE/CVF Conference on
               Computer Vision and Pattern Recognition},
  year      = {2023}
}

@article{smith2011network,
  title={Network modelling methods for FMRI},
  author={Smith, Stephen M and Miller, Karla L and Salimi-Khorshidi, Gholamreza and Webster, Matthew and Beckmann, Christian F and Nichols, Thomas E and Ramsey, Joseph D and Woolrich, Mark W},
  journal={Neuroimage},
  volume={54},
  number={2},
  pages={875--891},
  year={2011},
  publisher={Elsevier}
}

@article{huth2012continuous,
  title={A continuous semantic space describes the representation of thousands of object and action categories across the human brain},
  author={Huth, Alexander G and Nishimoto, Shinji and Vu, An T and Gallant, Jack L},
  journal={Neuron},
  volume={76},
  number={6},
  pages={1210--1224},
  year={2012},
  doi={10.1016/j.neuron.2012.10.014}
}

@article{deheer2017hierarchical,
  title={The hierarchical cortical organization of human speech processing},
  author={de Heer, Wendy A and Huth, Alexander G and Griffiths, Thomas L and Gallant, Jack L and Theunissen, Fr{\'e}d{\'e}ric E},
  journal={Journal of Neuroscience},
  volume={37},
  number={27},
  pages={6539--6557},
  year={2017},
  doi={10.1523/JNEUROSCI.3267-16.2017}
}

@article{tavor2016task,
  title={Task-free MRI predicts individual differences in brain activity during task performance},
  author={Tavor, Ido and Jones, O Parker and Mars, Rogier B and Smith, Stephen M and Behrens, Timothy E and Jbabdi, Saad},
  journal={Science},
  volume={352},
  number={6282},
  pages={216--220},
  year={2016},
  publisher={American Association for the Advancement of Science}
}

@article{fox2006coherent,
  title={Coherent spontaneous activity accounts for trial-to-trial variability in human evoked brain responses},
  author={Fox, Michael D and Snyder, Abraham Z and Zacks, Jeffrey M and Raichle, Marcus E},
  journal={Nature neuroscience},
  volume={9},
  number={1},
  pages={23--25},
  year={2006},
  publisher={Nature Publishing Group UK London}
}

@article{hasson2008hierarchy,
	author = {Hasson, Uri and Yang, Eunice and Vallines, Ignacio and Heeger, David J. and Rubin, Nava},
	title = {A Hierarchy of Temporal Receptive Windows in Human Cortex},
	volume = {28},
	number = {10},
	pages = {2539--2550},
	year = {2008},
	doi = {10.1523/JNEUROSCI.5487-07.2008},
	publisher = {Society for Neuroscience},
	issn = {0270-6474},
	URL = {https://www.jneurosci.org/content/28/10/2539},
	eprint = {https://www.jneurosci.org/content/28/10/2539.full.pdf},
	journal = {Journal of Neuroscience}
}

@article{hasson2004intersubject,
  author = {Uri Hasson and Yuval Nir and Ifat Levy and Galit Fuhrmann and Rafael Malach},
  title = {Intersubject synchronization of cortical activity during natural vision},
  year = {2004},
  journal = {Science},
  volume = {303},
  chapter = {1634-1640},
  doi = {10.1126/science.1089506},
}

@article{friston2000nonlinear,
  author = {Friston, Karl J. and Mechelli, Andrea and Turner, Robert and Price, Cathy J.},
  title = {Nonlinear Responses in {fMRI}: The Balloon Model, Volterra Kernels, and Other Hemodynamics},
  journal = {NeuroImage}, volume = {12}, number = {4}, pages = {466--477}, year = {2000},
  doi = {10.1006/nimg.2000.0630}
}

@article{assran2025vjepa2, author={Assran, Mido and others}, title={{V-JEPA 2}: Self-Supervised Video Models Enable Understanding, Prediction and Planning}, journal={arXiv preprint arXiv:2506.09985}, year={2025}}

@article{dascoli2025tribe, author={d'Ascoli, St{\'e}phane and Rapin, J{\'e}r{\'e}my and Benchetrit, Yohann and Banville, Hubert and King, Jean-R{\'e}mi}, title={{TRIBE}: {TRImodal Brain Encoder} for Whole-Brain {fMRI} Response Prediction}, journal={arXiv preprint arXiv:2507.22229}, year={2025}}

@article{dascoli2026foundation, author={d'Ascoli, St{\'e}phane and Rapin, J{\'e}r{\'e}my and Benchetrit, Yohann and Brooks, Teon and Begany, Katelyn and Raugel, Jos{\'e}phine and Banville, Hubert and King, Jean-R{\'e}mi}, title={A Foundation Model of Vision, Audition, and Language for In-Silico Neuroscience}, journal={arXiv preprint arXiv:2605.04326}, year={2026}}

@article{dong2026neuroworld, author={Dong, Zijian and Zhou, Jianxiong and Ng, Kwun Kei and Balagtas, Jan Paolo Macapinlac and Li, Zhizhou and Chen, Zijiao and Zhou, Juan Helen}, title={{NeuroWorld}: A Latent Brain World Model for Stimulus-Conditioned Human Brain Dynamics}, journal={arXiv preprint arXiv:2608.01773}, year={2026}}

@article{finn2020idiosynchrony, author={Finn, Emily S. and Glerean, Enrico and Khojandi, Arman Y. and Nielson, Dylan and Molfese, Peter J. and Handwerker, Daniel A. and Bandettini, Peter A.}, title={Idiosynchrony: From Shared Responses to Individual Differences during Naturalistic Neuroimaging}, journal={NeuroImage}, volume={215}, pages={116828}, year={2020}}

@article{gifford2025algonauts, author={Gifford, Alessandro Thomas and others}, title={The {Algonauts} Project 2025 Challenge: How the Human Brain Makes Sense of Multimodal Movies}, journal={arXiv preprint arXiv:2501.00504}, year={2025}}

@article{gokce2026mirage, author={Gokce, Abdulkadir and AlKhamissi, Badr and Schrimpf, Martin}, title={{MIRAGE}: Adaptive Multimodal Gating for Whole-Brain {fMRI} Encoding}, journal={arXiv preprint arXiv:2605.29850}, year={2026}}

@misc{hcp7tprotocol, author={{Human Connectome Project}}, title={{HCP} Protocols: Young Adult 7T Imaging}, year={2026}, note={Accessed September 2026}}

@article{lahner2024boldmoments, author={Lahner, Benjamin and Dwivedi, Kshitij and Iamshchinina, Polina and others}, title={Modeling Short Visual Events through the {BOLD} Moments Video {fMRI} Dataset and Metadata}, journal={Nature Communications}, volume={15}, pages={6241}, year={2024}}

@article{logothetis2001neurophysiological, author={Logothetis, Nikos K. and Pauls, Jon and Augath, Mark and Trinath, Torsten and Oeltermann, Axel}, title={Neurophysiological Investigation of the Basis of the {fMRI} Signal}, journal={Nature}, volume={412}, pages={150--157}, year={2001}}

@inproceedings{mai2025synbrain, author={Mai, Weijian and others}, title={{SynBrain}: Enhancing Visual-to-{fMRI} Synthesis via Probabilistic Representation Learning}, booktitle={Advances in Neural Information Processing Systems}, year={2025}}

@misc{meta2024llama32, author={{Meta AI}}, title={{Llama-3.2-3B}: Model Card}, year={2024}, url={https://huggingface.co/meta-llama/Llama-3.2-3B}}

@article{mueller2013variability, author={Mueller, Sophia and Wang, Danhong and Fox, Michael D. and Yeo, B. T. Thomas and Sepulcre, Jorge and Sabuncu, Mert R. and Shafee, Rebecca and Lu, Jie and Liu, Hesheng}, title={Individual Variability in Functional Connectivity Architecture of the Human Brain}, journal={Neuron}, volume={77}, number={3}, pages={586--595}, year={2013}, doi={10.1016/j.neuron.2012.12.028}}

@article{nozari2024macroscopic, author={Nozari, Erfan and Bertolero, Maxwell A. and Stiso, Jennifer and others}, title={Macroscopic Resting-State Brain Dynamics Are Best Described by Linear Models}, journal={Nature Biomedical Engineering}, volume={8}, pages={68--84}, year={2024}}

@article{paugam2024benchmark, author={Paugam, Fran{\c c}ois and Pinsard, Basile and Lajoie, Guillaume and Bellec, Pierre}, title={A Benchmark of Individual Auto-Regressive Models in a Massive {fMRI} Dataset}, journal={Imaging Neuroscience}, volume={2}, pages={imag--2--00228}, year={2024}}

@article{schaefer2018localization, author={Schaefer, Alexander and Kong, Ru and Gordon, Evan M. and others}, title={Local-Global Parcellation of the Human Cerebral Cortex from Intrinsic Functional Connectivity {MRI}}, journal={Cerebral Cortex}, volume={28}, number={9}, pages={3095--3114}, year={2018}}

@article{seamless2023communication, author={{Seamless Communication} and others}, title={{Seamless}: Multilingual Expressive and Streaming Speech Translation}, journal={arXiv preprint arXiv:2312.05187}, year={2023}}

@article{wen2018neural, author={Wen, Haiguang and Shi, Junxing and Zhang, Yizhen and Lu, Kun-Han and Cao, Jiayue and Liu, Zhongming}, title={Neural Encoding and Decoding with Deep Learning for Dynamic Natural Vision}, journal={Cerebral Cortex}, volume={28}, number={12}, pages={4136--4160}, year={2018}}

@article{wen2018transferring, author={Wen, Haiguang and Shi, Junxing and Chen, Wei and Liu, Zhongming}, title={Transferring and Generalizing Deep-Learning-Based Neural Encoding Models across Subjects}, journal={NeuroImage}, volume={176}, pages={152--163}, year={2018}}

@article{yeshurun2017same, author={Yeshurun, Yaara and Swanson, Stephen and Simony, Erez and Chen, Janice and Lazaridi, Christina and Honey, Christopher J. and Hasson, Uri}, title={Same Story, Different Story: The Neural Representation of Interpretive Frameworks}, journal={Psychological Science}, volume={28}, number={3}, pages={307--319}, year={2017}, doi={10.1177/0956797616682029}}

@article{yin2026brainvista, author={Yin, Xuanhua and Zhao, Runkai and Yao, Lina and Cai, Weidong}, title={{BrainVista}: Modeling Naturalistic Brain Dynamics as Multimodal Next-Token Prediction}, journal={arXiv preprint arXiv:2602.04512}, year={2026}}
\bibliographystyle{unsrtnat}

\clearpage
\appendix
\raggedbottom
\startcontents[appendices]
\section*{Appendix Contents}
\printcontents[appendices]{}{1}{\setcounter{tocdepth}{2}}
\clearpage

\section{Dataset and Protocol Details}
\label{app:data}
\begin{table}[htbp]
\centering
\small
\caption{
Dataset designs and cross-subject splits.
A stimulus unit denotes a unique movie clip for HCP and BOLD Moments,
and a unique movie segment for Algonauts.
Unique duration excludes fixation and repeated presentations.
BOLD Moments additionally includes 4-s null trials.
Full details are provided in Appendix~\ref{app:data}.
}
\label{tab:datasets}

\setlength{\tabcolsep}{6pt}
\renewcommand{\arraystretch}{1.05}
\begin{tabular}{@{}lccc@{}}
\toprule
& HCP four movies & BOLD Moments & Algonauts 2025 \\
\midrule
Subjects
& 154 & 10 & 4 \\

Subject split
& 107/15/32 & Five two-subject folds & Four LOSO folds \\

TR
& 1.00 s & 1.75 s & 1.49 s \\

Movie duration
& 49.4 min & 5.1 min & 63.1 h \\

Unique clips
& 15 & 102 & 331 \\

Clip duration
& 1.1--4.3 min & 3 s & $\sim$10--12 min \\

Between-clip interval
& 20 s fixation & 1 s fixation & Between-run break \\

Clips / run
& 4--5
& 85
& 1 \\

Train--test stimulus overlap
& 100\%
& 100\%
& 89.7\% \\

Scored prediction window
& (1.1--4.3 min) / clip
& (7.5 min) / run
& (9.0--14.4 min) / clip \\

\bottomrule
\end{tabular}
\end{table}
\subsection{HCP four movies}
\label{app:hcp_fourmovie_new}

We use four movie-viewing scans from each of 154 HCP subjects,
giving 616 scan recordings. The four scans contain 5, 4, 5, and 4
clip presentations, respectively. These 18 presentations correspond
to 15 unique clips because some content is repeated.
The analysis therefore contains 2,772 clip presentations, but repeated
presentations are not counted as distinct stimuli.
When movie versions differ across subjects, stimulus features
correspond to the version actually presented to each subject.

Subjects are partitioned into 107 training, 15 validation, and 32 test
subjects, yielding 1,926, 270, and 576 clip records, respectively.
Each clip presentation is treated as a separate prediction and scoring
unit.
The preceding fixation supplies the initial measured BOLD history,
and every BOLD sample within the clip is included as a prediction target.
Prediction blocks never cross clip boundaries.

We extract regional time series using the 1,000-parcel, seven-network
Schaefer atlas. The 2-mm atlas label image is resampled onto the
1.6-mm BOLD sampling grid using nearest-neighbor interpolation,
which preserves discrete parcel labels.
Voxel signals are then averaged within each parcel.
We use the supplied hp2000-clean BOLD data without applying an
additional temporal filter or per-record target normalization.
Stimulus features sampled at 2 Hz are averaged in consecutive pairs
to match the one-second BOLD grid.
The stimulus-to-BOLD delay and training-set normalization are described
in Appendix~\ref{app:stimulus_delay}.

\subsection{BOLD Moments}
\label{app:bold_preprocessing}

We use the version-B, fMRI Prep-preprocessed volumetric BOLD data
from ten subjects in MNI152\-NLin2009c\-Asym space.
Our analysis uses the continuous test-task scans from sessions
02--05, with three scans per session, giving 120 scan recordings.
These scans represent 102 unique video clips, each lasting 3 seconds,
with 1-second fixation intervals and additional 4-second null trials.
The prediction targets are continuous BOLD time series.

Because stimulus order differs across subjects, each scan is aligned
using its own event identities and onset times.
An entire scan is treated as one prediction record.
A prediction block may therefore include several stimulus events,
but it cannot cross a scan boundary.
The evaluation measures generalization to unseen subjects, rather
than generalization to a held-out set of video identities.

Regional signals are obtained by resampling the FSLMNI152
Schaefer-1000 label image onto each BOLD grid with nearest-neighbor
interpolation and averaging voxel signals within parcels.
This operation aligns the sampling grids using their spatial
coordinates.

Nuisance regression is performed separately for each scan.
Available regressors include the six motion parameters, their temporal
derivatives, their squares and squared derivatives, white-matter and
CSF signals, cosine regressors, and non-steady-state indicators.
An intercept and linear trend are also included, and missing confound
values are interpolated.
Ordinary least squares residuals are centered and divided by their
sample standard deviation within each complete scan.
This offline preprocessing is applied to training, validation, and
test scans alike, including the full duration of held-out scans.

Video features are aligned to scanner sampling times relative to each
event onset using interpolation of the 2-Hz feature sequence.
Queries outside a clip's available feature times use its nearest
endpoint value.
Intervals without represented stimulus content retain their BOLD
targets, with unavailable input modalities marked as missing.
In particular, oddball events without corresponding movie features
are treated as unmodeled inputs, not as known fixation.

Subjects viewed silent videos.
Audio features and ASR-derived text, when included in a multimodal
comparison, are extracted from the original video soundtracks.
They are therefore auxiliary descriptions of stimulus content,
not auditory or linguistic signals actually presented to subjects.
ASR text is not a visual description of the video.

Applying the three-sample stimulus-to-BOLD delay to each 268-TR scan
leaves 265 aligned samples.
The first eight aligned BOLD samples provide the initial history,
and the remaining 257 samples are scored at every refresh interval.
The five test-subject pairs are (01,06), (02,07), (03,08),
(04,09), and (05,10).
For each fold, the other eight subjects each contribute eleven
training scans and one validation scan, namely the third scan of
session 05.
Each fold thus contains 88 training, eight validation, and 24 test
records.
No test subject contributes to predictor fitting or to the shared
training-set normalization statistics.

\subsection{Algonauts 2025}

We use the publicly labeled recordings from subjects 01, 02, 03,
and 05, retaining the native 1,000-parcel targets and a TR of
1.49 seconds.
Each subject contributes 348 recorded segments, giving 1,392
records across subjects and 331 unique stimulus feature timelines.
Each recording is processed and evaluated separately.

We construct four leave-one-subject-out folds.
In each fold, all 348 records of one subject are reserved for testing,
while the remaining three subjects contribute 927 training and
117 validation records.
Training and validation are partitioned by grouped stimulus
identities, using Friends episode and movie groups, rather than
independently assigning temporal windows.
This partition is selected once with split seed 0 and is shared
by all models and training seeds.
These folds define our cross-subject benchmark and should not be
interpreted as a reproduction of the original challenge evaluation
protocol.

Stimulus features follow the released event-based temporal alignment:
feature samples assigned to the same rounded time bin are averaged.
The aligned data already associate each BOLD sample with stimulus
features three TRs earlier, corresponding to 4.47 seconds, so no
additional temporal shift is applied.
We retain the source whole-record standardization and subsequently
apply the shared training-set normalization described below.
The first eight aligned BOLD samples initialize prediction, and
all remaining samples are scored without crossing recording
boundaries.

\subsection{Stimulus representations and temporal alignment}\label{app:stimulus_delay}

The primary multimodal comparisons use frozen Llama 3.2-3B,
Wav2Vec-BERT 2.0, and V-JEPA2 ViT-g extractors for text, audio,
and video, respectively.
Two group-averaged representations from each extractor yield
feature shapes of \(2\times3072\), \(2\times1024\), and
\(2\times1408\).
Flattening and concatenating these representations gives 11,008
input dimensions: 6,144 for text, 2,048 for audio, and 2,816 for
video.
ASR and feature extraction are performed before predictor fitting,
and the stimulus extractors are not fine-tuned.

Stimulus representations are first aligned to each dataset's BOLD
sampling grid using the dataset-specific procedures above.
We then pair a BOLD observation at sample \(t\) with the stimulus
representation at sample \(t-3\):
\[
    \widetilde{\mathbf{x}}_t = \mathbf{x}_{t-3}.
\]
The shift is applied once, whether it is already present in the
released aligned data or introduced during our preprocessing.
It corresponds to 3.00 seconds for HCP, 5.25 seconds for BOLD
Moments, and 4.47 seconds for Algonauts.

All models receive stimulus features aligned to the same target
times, although their temporal receptive fields within the supplied
stimulus interval may differ.
Stimulus extraction is offline and is not required to be causal
with respect to the stimulus timeline.

Records with missing modalities are retained.
Unavailable or experimentally ablated modalities are set to zero
after normalization.
A modality-presence mask determines which entries participate in
normalization and which are zeroed, but the mask is not supplied
as an additional learned input channel.

\subsection{Normalization and information separation}

For each fold, we estimate feature means and population standard
deviations from available training features only.
An identical stimulus feature sequence used by multiple subjects
is counted once when estimating these statistics.
Target means and population standard deviations are estimated
from the training BOLD records.
Normalization is performed independently for each feature
dimension and each target ROI, with standard deviations floored
at \(10^{-4}\).

The resulting transformation is shared by all models within a
fold, including TRIBE v2 trained from scratch, and is applied
unchanged to validation and test data.
Validation data are used for model selection, while test data
are used only for final evaluation.
Neither validation nor test observations contribute to the
shared normalization statistics.

This training-set transformation is separate from dataset-level
offline preprocessing.
In particular, the complete-scan nuisance regression and
standardization used for BOLD Moments, and the retained source
standardization for Algonauts, precede predictor fitting and
also operate on held-out recordings.
Consequently, the benchmark evaluates prediction of offline
preprocessed BOLD; it is not an end-to-end online preprocessing
benchmark.
Within this setting, predictors receive measured BOLD only
according to the refresh protocol below.

\subsection{Prediction protocol}
\label{app:prediction_protocol}
We distinguish three temporal quantities. The context length $C$ is the number of measured BOLD samples used to initialize each prediction block. The training rollout length $K$ is the number of targets in a training prediction window. The test refresh interval $H$ is the maximum number of consecutive targets predicted before measured BOLD is supplied again. For history-conditioned predictors at finite $H$, the target interval of each recording is divided into consecutive prediction blocks of up to $H$ frames (Figure~\ref{fig:rollout_protocol}). Each block starts from the latest $C$ measured BOLD frames preceding it and generates predictions autoregressively. The predicted blocks are concatenated within each recording before temporal correlation is computed against the corresponding ground truth. We use $C=8$ for all datasets, with $K=64$ for HCP and BOLD Moments and $K=92$ for Algonauts.

We evaluate the finite refresh intervals
\[
H\in\{1,2,4,8,16,32,48,64,80,96,128,160,192\}.
\]
These intervals are expressed in samples on the dataset-specific
TR grids listed in Table~\ref{tab:datasets}.
Predictions are assembled in temporal order and cover the same
target samples for every \(H\), including a final block shorter
than \(H\) when necessary.
No prediction block crosses a recording boundary.
History-free predictors instead use fixed \(K\)-step stimulus
windows independent of \(H\); their predictions therefore do
not change with the measured-BOLD refresh interval.

\begin{figure}[H]
\centering
\includegraphics[width=0.98\linewidth]{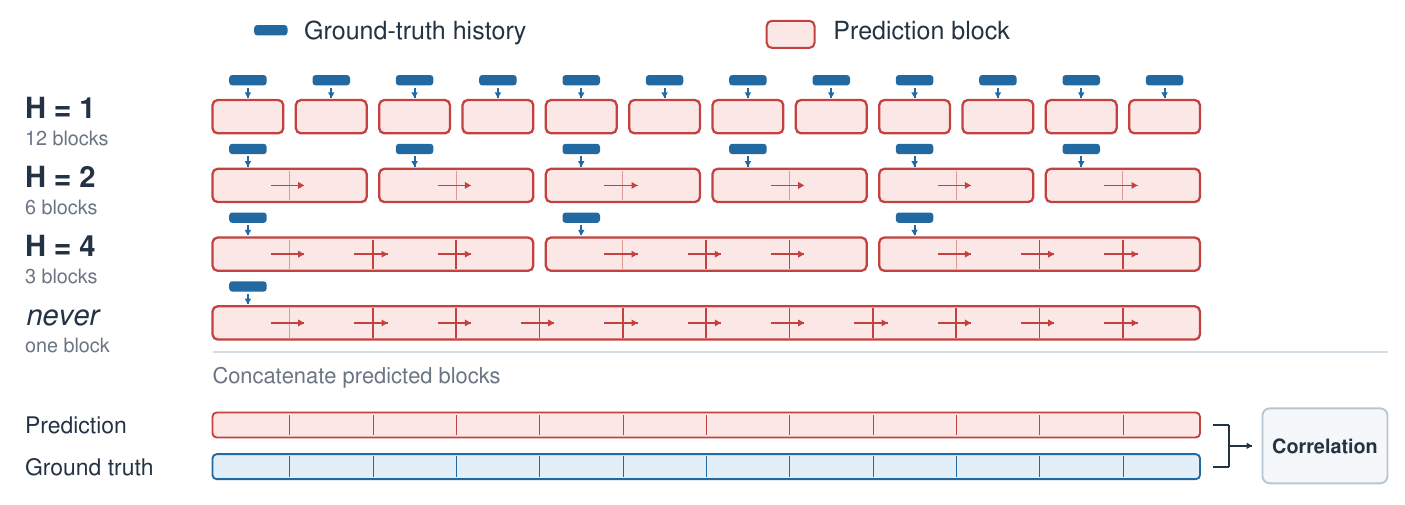}

\caption{{Prediction blocks for \(H=1,2,4\) and \emph{never}, illustrated for 12 target frames. Blue markers indicate the measured BOLD history supplied at each block start; supplying measured history again between blocks is termed a \emph{refresh}. Arrows within red blocks indicate autoregressive rollout. Predicted blocks are concatenated before scoring.}}
\label{fig:rollout_protocol}
\end{figure}

\paragraph{No-refresh evaluation.}
\label{app:extended_snapshots}

The \emph{never} condition supplies only the initial measured
BOLD context of each recording and performs no subsequent
refresh.
Its prediction duration depends on the finite recording length;
it is not an infinite-time limit.
Finite-\(H\) and no-refresh evaluations use the same selected
model parameters and score the same target samples.
Changing \(H\) does not involve retraining.

\subsection{Metrics, aggregation, and uncertainty}
\label{app:benchmark_evaluation}

Prediction accuracy is measured by temporal Pearson correlation.
For subject \(s\), recording \(c\), ROI \(i\), and refresh interval
\(H\), we correlate the predicted and observed time series over
the scored target samples \(\mathcal{T}_c\):
\[
r_{s,c,i}^{(H)}
=
\operatorname{corr}_{t\in\mathcal{T}_c}
\left(
y_{s,c,t,i},
\widehat{y}_{s,c,t,i}^{(H)}
\right).
\]

We first average valid ROI correlations within each recording,
then average recordings within each subject, and finally average
test subjects with equal weight:
\[
r^{(H)}
=
\frac{1}{|\mathcal{S}_{\mathrm{test}}|}
\sum_{s\in\mathcal{S}_{\mathrm{test}}}
\frac{1}{|\mathcal{C}_s|}
\sum_{c\in\mathcal{C}_s}
\frac{1}{|\mathcal{R}_{s,c}|}
\sum_{i\in\mathcal{R}_{s,c}}
r_{s,c,i}^{(H)}.
\]
Here, \(\mathcal{C}_s\) is the set of evaluated recordings for
subject \(s\), and \(\mathcal{R}_{s,c}\) is the set of valid
target ROIs for that recording.
Recordings receive equal weight within a subject even when
their durations differ; longer clips do not receive additional
weight merely because they contain more samples.
We neither concatenate different clips to calculate a single
correlation nor substitute spatial correlation across ROIs for
temporal correlation.

ROIs with constant target time series are excluded using the
same target-derived mask for every model.
A constant prediction for a nonconstant target is assigned a
correlation of zero.
Nonfinite predictions are treated as evaluation failures rather
than silently omitted.

\paragraph{Seed averaging and between-subject variability.}
Each designated test subject appears in exactly one test fold
per seed, and dataset-level results include all designated folds.
For each neural model, we first compute a subject-level score
for each training seed using the ROI and recording aggregation
described above, then average these scores across seeds 0, 1, and 2.
The reported dataset mean is the equally weighted average of
these seed-averaged subject scores.
Error bars show the sample standard deviation across held-out
subjects after seed averaging, describing between-subject
variability rather than variability across training seeds or
a confidence interval for the population mean.
Ridge models contribute one deterministic score per subject;
their repeated evaluations across seeds are not treated as
independent observations.
All comparisons use the same designated test subjects and,
for neural models, the complete seed set.

Regional comparisons likewise include all designated test
subjects and neural seeds 0, 1, and 2.
Model differences are paired within seed before averaging,
while absolute ridge profiles reuse the deterministic fits.
No ROI is selected on the basis of its test-set prediction gain.

\section{Implementation Details}
\label{app:implementation}

\subsection{Model taxonomy and benchmark scope}
\label{app:model_taxonomy}

Table~\ref{tab:related_model_classes} summarizes representative prediction approaches by their inputs and downstream architectures. It describes the cited methods; the benchmark adaptations, diagnostic references, and training procedures are specified below.

\begin{table}[h]
\centering
\scriptsize
\setlength{\tabcolsep}{4pt}
\renewcommand{\arraystretch}{1.1}
\caption{Representative downstream prediction approaches, organized by input and prediction algorithm. Linear and nonlinear refer to the downstream predictor, independently of the stimulus encoder. Benchmark adaptations and diagnostic references are described in Section~\ref{sec:baselines}.}
\label{tab:related_model_classes}
\begin{tabular}{@{}p{0.19\linewidth}p{0.12\linewidth}p{0.32\linewidth}p{0.30\linewidth}@{}}
\toprule
\textbf{Predictor input} & \textbf{Mapping} & \textbf{Prediction algorithm} & \textbf{Representative work} \\
\midrule
Stimulus features & Linear & Ridge-regression voxel-wise readout & Wen et al.~\citep{wen2018neural} \\
\addlinespace
Stimulus features & Nonlinear & Temporal Transformer & TRIBE; TRIBE v2 \citep{dascoli2025tribe,dascoli2026foundation} \\
\addlinespace
Stimulus features & Nonlinear & Adaptive layer aggregation and temporal brain encoder & MIRAGE  \citep{gokce2026mirage} \\
\addlinespace
BOLD history & Linear & Linear autoregressive models; linearized graph-based predictors & Nozari et al.; Paugam et al.~\citep{nozari2024macroscopic,paugam2024benchmark} \\
\addlinespace
Stimulus features and BOLD history & Nonlinear & Brain tokenizer and autoregressive Transformer & BrainVista \citep{yin2026brainvista} \\
\addlinespace
Stimulus features and BOLD history & Nonlinear & Latent dynamics model and rollout decoder & NeuroWorld \citep{dong2026neuroworld} \\
\bottomrule
\end{tabular}
\end{table}

\subsection{Configuration, model classes, and sampling}

The benchmark uses \(C=8\) measured context frames and training/validation rollout length \(K=64\) on HCP and BOLD Moments or \(K=92\) on Algonauts. These window lengths are distinct from the test refresh interval \(H\). {{For neural models, \(K\) specifies the prediction-window length used for training and validation-based checkpoint selection. Ridge coefficients are fitted from all available one-step training rows, and penalties are selected using validation BOLD MSE. For RR-AR, validation uses autonomous \(K\)-step rollouts.}} Each training record contributes \(\lceil(T-C)/K\rceil\) sampled windows per epoch. Starts are drawn uniformly from legal C+K windows using a generator keyed by training seed, epoch, and sample index. Therefore different models have the same candidate record/start identities at the same seed and epoch even when batch sizes differ. Windows never cross a record boundary. Hidden state is reset per window; the true prefix warms the recurrent state, and no true BOLD is injected within its K-step prediction.

Validation replaces random starts with the deterministic partition
\[
a_j=C+jK,\qquad b_j=\min(T,a_j+K),\qquad a_j<T.
\]
Block j receives true BOLD \([a_j-C,a_j)\) and predicts \([a_j,b_j)\). All target indices C through T-1 occur exactly once. Variable-length tails are evaluated at their true length and grouped into homogeneous sub-batches, with no padded attention tokens or discarded targets. Validation is global squared-error sum divided by the global number of scalar targets, not the mean of per-batch means.

Validation runs every epoch. A strict loss improvement saves the checkpoint; ties keep the earlier checkpoint. Patience is 10 unsuccessful validations after the native warmup; the counter restarts at each stage. An epoch-1 checkpoint remains eligible even during warmup. The best model is restored before testing or the next stage. Native stage ceilings remain in effect. No old-protocol checkpoints are imported.

\subsection{Optimization settings and configuration sources}
\label{app:bold_baseline_details}

\begin{table}[htbp]
\centering
\scriptsize \caption{Declared optimization ceilings and schedules. Every neural stage has patience 10; final prediction stages select by complete-coverage BOLD MSE. Pretraining uses the indicated stage loss. Numbers are not a claim that every stage runs to its ceiling.}
\label{tab:bold_baseline_training}
\resizebox{\linewidth}{!}{%
\begin{tabular}{lllllll}
\toprule
Model/stage & Max epochs & Batch & Optimizer & LR & WD & Schedule / clip norm\\
\midrule
BrainVista tokenizer & 50 & 32 & AdamW & $2\cdot10^{-5}$ & .01 & Constant / .5\\
BrainVista predictor & 40 & 32 & AdamW & $2\cdot10^{-4}$ & .01 & 5\% warmup, cosine / .5\\
NeuroWorld LDL & 50 & 16 & AdamW & $2\cdot10^{-4}$ & .01 & 100-update warmup, cosine / .5\\
NeuroWorld decoder & 200 & 16 & AdamW & $10^{-3}$ & .01 & 500-update warmup, cosine / .5\\
TRIBE v2 scratch & 15 & 8 & Adam & $10^{-4}$ & 0 & 10\% warmup, OneCycle / none\\
MIRAGE & 15 & 16 & AdamW & $10^{-4}$ & .01 & 10\% warmup, OneCycle / 1.0\\
Ridge models & Closed form & All training rows & Eigendecomposition & -- & -- & Validation-selected $\lambda$\\
\bottomrule
\end{tabular}}
\end{table}

All Adam-family optimizers use \(\beta=(0.9,0.999)\) and \(\epsilon=10^{-8}\). BrainVista excludes bias and one-dimensional normalization parameters from weight decay. Neural operations use bf16 autocast, except for the explicitly documented NeuroWorld FP32 recovery fits in Appendix~\ref{app:cross_model_inputs_completed}; Main-benchmark ridge fitting uses float64, with coefficients and intercepts cast to float32 for prediction; exceptions for representation controls are specified in Appendix~\ref{app:visual_representation_all}. TF32 is disabled; cuDNN deterministic mode is enabled and benchmarking disabled. Dataset-specific loader workers are 2/4/2; this is an I/O setting, not a different sampling or evaluation rule.

\begin{table}[htbp]
\centering
\scriptsize \caption{Configuration provenance and deliberate comparison adaptations. Scientific choices are declared in the shared configuration and, for MIRAGE, its explicit adapter settings, not runtime paper defaults.}
\label{tab:config_sources}
\begin{tabular}{lp{0.72\linewidth}}
\toprule
Component & Source and scope\\
\midrule
Shared input/test protocol & Study design: C=8, dataset K, HRF=3 TR,
fixed three-modality slots, full validation MSE, patience10 and temporal ROI corr.
These replace model-native evaluation conventions.\\
BrainVista & Published two-stage structure and nominal training budgets;
our implementation supplies dimension reconciliation, mask/filter choices,
atlas-label grouping and shared autonomous BOLD-MSE training.\\
NeuroWorld & Published LDL/decoder structure and nominal budgets;
latent widths, SIGReg settings and shared decoder are specified implementation
choices; the adapter supplies all C observed frames.\\
TRIBE v2 & Released FmriEncoder class and architecture/config-derived optimizer
schedule; scratch initialization, atlas-sized output, shared normalization,
common input interval and val-loss selection are comparison adaptations.\\
Ridge & Explicit affine design, covariance divided by training-row count,
intercept unpenalized; adaptive boundary search uses common K-step val MSE.\\
MIRAGE & Released downstream code and declared width/depth;
common cached features, shared ROI readout, no test-subject head fitting,
and common validation-MSE selection.\\
\bottomrule
\end{tabular}
\end{table}

\paragraph{MIRAGE architecture and scope.}
\label{app:mirage_details}
We use the released downstream model classes with hidden width 3,072, eight Transformer layers, eight heads and fourfold feed-forward expansion. The layer-gating module uses four heads, 24 queries and attention dropout 0.2; modality dropout is 0.3. The Transformer attention, feed-forward and layer dropout settings are zero. Linear stimulus projectors and the upstream group-linear ROI head are retained. Allocated downstream sizes for the current atlas are reported in Table~\ref{tab:parameter_counts}.

This is a matched-feature downstream adaptation, not a reproduction of native pretrained MIRAGE scores. The inputs are the common two-group Llama/w2v-BERT/V-JEPA cached features, rather than full-layer Qwen-Omni representations. Subject embeddings are disabled and one shared ROI head is trained on the training subjects; no unseen-subject head is fitted. Temporal output pooling is disabled to preserve the aligned TR grid. Bidirectional stimulus attention is retained within fixed C+K tiles; no measured BOLD vector enters the model. The same fixed tiles are used for every H, including never. All downstream weights are initialized from scratch at seeds 0/1/2 and trained using the shared sampler, AdamW, 15-epoch ceiling, batch size 16, learning rate $10^{-4}$, weight decay 0.01, OneCycle schedule with 10\% warmup and gradient clipping at 1.0. Checkpoint selection uses complete-coverage validation BOLD MSE, not the upstream validation-Pearson criterion. The common BF16/TF32-disabled policy and all held-out subjects are retained. These adaptations isolate the released downstream architecture under common inputs; they do not evaluate the complete native pretrained system.

\paragraph{BrainVista architecture and stages.}
Actual atlas network labels partition all ROI columns exactly once into seven groups; consecutive equal-size fallback groups are not used. Each network has a two-linear-layer GELU autoencoder with token width \(\max(1,\operatorname{round}(0.8n_g))\) and intermediate width \(\max(n_g,\mathrm{tokenwidth})\). The stitched tokens project to width 256. Depthwise brain/stimulus convolutions have widths 3/5 and padding 1/2. Token-type and learned position embeddings precede four pre-normalized Transformer layers, eight heads, a 1,024-wide GELU feed-forward layer and dropout 0.1. A four-head spatial mixer operates at width 256. The model predicts a token residual added to the last token.

The configured attention-mask mode is named \texttt{released} in our code; this name is not a claim of complete official-code reproduction. It imposes causal token order and blocks brain-query/brain-key attention, while stimulus queries may attend earlier brain keys. Temporal convolutions stay within the supplied history/prediction window. All eight observed brain frames are tokenized. Generated latent tokens are recursively appended; the active attention window retains C tokens.

Tokenizer training minimizes reconstruction MSE on the C+K measured frames. The best tokenizer is frozen for predictor training. Predictor training differentiates through autonomous K-step latent rollout and the frozen decoder, using decoded BOLD MSE rather than the legacy teacher-forced latent-token loss. No test-subject decoder is fitted.

\paragraph{NeuroWorld architecture and stages.}
Atlas-network MLPs concatenate into a LayerNorm-normalized 256-dimensional latent; widths are split across the seven networks. A Linear--GELU--LayerNorm stimulus encoder maps 11,008 features to 256. Four causal adaptive-LayerNorm transition blocks use eight heads, fourfold feed-forward expansion, and dropout 0.1. The decoder is a \(256\to256\) Linear--GELU--LayerNorm trunk and a \(256\to R\) linear head. Despite its internal name \texttt{subject\_head}, this adapter uses one shared head, not a fitted head for each unseen test subject.

LDL learns next-latent prediction over the K target transitions with teacher forcing and SIGReg weight 0.01. SIGReg uses 256 fixed normalized random directions with projection seed1701. This is a latent-pretraining objective, not a reported BOLD metric. The best world model is then frozen in evaluation mode, including its dropout, while the decoder is trained on autonomous K-step latent rollouts by BOLD MSE. Rollout starts from the encoded sequence of all C observations and retains generated latent history; it does not initialize from only the last measured frame.

\paragraph{TRIBE v2.}
The released FmriEncoder uses hidden width 1,152, eight Transformer layers, eight heads and fourfold feed-forward expansion. The configuration retains rotary positions, time-position embeddings, ScaleNorm and residual scaling, with maximum sequence length 1,024 and low-rank readout width 2,048. Attention is bidirectional within the supplied stimulus interval, not over an entire scanner run. Modality dropout is 0.3, subject dropout 0.1; ordinary, attention, feed-forward and temporal dropout are zero.

Fixed-slot grouped features pass through the released projection and subject-conditioned readout. Training uses the training subject's head; validation/testing use the average training heads. No held-out subject embedding or readout is fitted. All predictor weights start from scratch; frozen feature extractors remain pretrained. OneCycle uses division factors 25 and 10,000 and momentum limits 0.85/0.95. Its official-style 15-epoch ceiling is retained, but checkpoint selection is common validation MSE, not the native validation Pearson criterion. At test, fixed K-step stimulus tiles are independent of H. Bidirectional access within a known stimulus tile is distinguished from access to future measured BOLD, which is never supplied.

\paragraph{Ridge design, solver, and selected regularization.}
With normalized aligned features \(\mathbf x_t\), the designs are
\[
\mathbf z_t^{\rm stim}=\mathbf x_t,\qquad
\mathbf z_t^{\rm AR}
=[\mathbf y_{t-8};\ldots;\mathbf y_{t-1};\mathbf x_t].
\]
All ROI lags can affect all predicted ROIs. Every training row with a complete eight-frame history is used; one-step fitting is closed form, not BPTT. Let \(Z\in\R^{N\times d}\) contain the centered training design rows, where \(d=CR+D\), and \(Y\in\R^{N\times R}\) the centered next-step targets. With this oldest-to-newest history ordering, $W=[A_C\;\cdots\;A_1\;B]^\top$, matching Equation~\ref{eq:rr}; the column-vector prediction is $\mathbf b+W^\top\mathbf z_t^{\rm AR}$. Writing \(G=Z^\top Z/N\) and \(Q=Z^\top Y/N\), the ridge estimator is
\begin{equation}
W_\lambda=\arg\min_W N^{-1}\|Y-ZW\|_F^2+\lambda\|W\|_F^2
 = (G+\lambda I)^{-1}Q.
\label{eq:ridge_fit}
\end{equation}
The unpenalized intercept is \(\mathbf b=\bar{\mathbf y}-W_\lambda^\top\bar{\mathbf z}\), where the training means are column vectors. At \(\lambda=0\), the numerical minimum-norm solution replaces the inverse. No further private feature scaling or stimulus-lag expansion is added. Training sufficient statistics are accumulated in chunks of 256 rows in float64. The centered Gram matrix is symmetrized and eigendecomposed in float64 to obtain the ridge solutions. The fitted coefficients and unpenalized intercept are cast to float32 for prediction. Candidate values begin at \(\{0.01,0.1,1,10,100\}\); a boundary optimum expands by a factor of ten in that direction, with a maximum of 12 expansions and an explicit failure if no interior optimum is found. Selection evaluates autonomous K-step validation MSE, not the one-step fitting residual or test correlation. {{Thus changing \(K\) leaves the candidate one-step fits unchanged but can change the selected \(\lambda\) and the final ridge predictor.}}

In BOLD fold0--4, stimulus Ridge selects \((10,10,10,10,10)\); RR-AR selects \((10,10,1,10,1)\). All are interior optima of the initial grid.

\subsection{Parameter-matched controlled transitions}
\label{app:controlled_transition}
The MLP and GRU receive the current stimulus vector and flattened C=8 measured/predicted BOLD history. The MLP uses an affine hidden layer, GELU and an affine ROI readout; the GRU uses a GRUCell and an affine ROI readout, resetting its hidden state at each measured-BOLD refresh. Both feed predicted BOLD back during rollout. Hidden widths match the dense RR coefficient budget, including offsets, within 1\%, without test selection. Neural weights use default random initialization, not RR weights or teachers.

Training retains the existing AdamW, K-step MSE, BF16 computation, 80-epoch budget and patience-10 full-validation protocol. Each prescribed learning rate ($10^{-4},3\times10^{-4},10^{-3}$) is assessed separately; minimum finite full-validation MSE selects the checkpoint. Numerical failure stops that candidate, retaining its best finite validated checkpoint and explicitly recording non-convergence; remaining candidates are still assessed. Test outcomes never trigger another checkpoint selection. Standard and extended evaluation use the same dataset-specific batch size and retain the H=96 replay check. All designated folds and seeds 0/1/2 are included in the main benchmark (Figure~\ref{fig:baseline}). Runs span the declared GPU/software
environments; this is not an identical-runtime comparison or a certificate
of globally optimal neural fitting.

\paragraph{Numerical handling.}
Seventeen endpoints encountered a nonfinite gradient or prediction in at
least one learning-rate candidate. The full prescribed candidate grid was
retained, and the validation-selected candidate in each case completed
normally. This denotes completion under the budget or early-stopping rule,
not certified convergence. Test scores were never used to reselect a candidate.
Evaluation replays used the same selected weights and a consistent batch size
for standard and extended refresh intervals; no retraining or test-based
checkpoint selection was introduced.

\subsection{Allocated parameter counts}
\label{app:parameter_counts}
\begin{table}[htbp]
\centering\small
\caption{Downstream model size in millions of allocated parameters or fitted coefficients, excluding frozen stimulus extractors. MLP and GRU are omitted from the table because their parameter counts match dense RR-AR within 1\% in each dataset. Rank64 counts the stored factors and output offsets of one prespecified reduced-rank setting; the full rank sweep is shown in Figure~\ref{fig:lowrank_absolute}.}
\label{tab:parameter_counts}
\begin{tabular}{lrrr}
\toprule
Model & HCP & BOLD & Algonauts\\
\midrule
RR-AR & 19.009 & 19.009 & 19.009\\
RR-AR (rank64) & 1.282 & 1.282 & 1.282\\
RR-stim & 11.009 & 11.009 & 11.009\\
TRIBE v2 scratch & 356.526 & 153.675 & 143.430\\
BrainVista & 7.649 & 7.649 & 7.649\\
NeuroWorld & 7.613 & 7.613 & 7.613\\
MIRAGE & 1097.679 & 1097.679 & 1097.679\\
\bottomrule
\end{tabular}
\end{table}

Counts include predictor parameters frozen between training stages and fitted ridge coefficients, but exclude all frozen feature extractors. Ridge has $(D+1)R$ stimulus-only coefficients including the intercept, or $(D+CR+1)R$ autoregressive coefficients. TRIBE counts depend on the number of training-subject readouts and atlas size. These counts do not imply equal effective capacity, training time, FLOPs, or memory consumption. The cross-model counts are reported in Table~\ref{tab:parameter_counts}. For rank \(r\), storing the factorized map requires \((D+CR+R)r+R\) coefficients, including the output offset. This formula applies to every prespecified rank.

\paragraph{Reproducibility.}
\label{app:provenance}
Training, validation selection and evaluation use the declared configurations
and folds. Tables and plots are aggregated from saved results, without changing
model weights or selecting by test performance. Reporting retains complete
declared fold/seed sets and distinguishes fixed-fold controls from all-fold
benchmarks. Source manifests, configurations and numerical failure records are
retained with the experiments.

\section{Stimulus Representation and Modality Controls}
\label{app:input_controls}
\label{app:cross_model_inputs_completed}
\label{app:modality_results}
Every reported neural input combination includes all designated held-out
subjects and seeds 0/1/2. Within each seed, records are averaged within subject
and subjects equally; error bars show the sample SD of the resulting seed means.
Ridge fits are deterministic. Each input combination is refitted and selected
using validation data, rather than masking a trained full-input model at test.
The same frozen feature slots, target grids and scoring samples are retained.
Full V/A/T inputs are reported in Figure~\ref{fig:baseline}; the
six reduced combinations and pixels are reported below.

BOLD Moments was presented silently. Its audio and ASR-text inputs describe
the unpresented source soundtrack and are auxiliary feature controls, not
evidence of sensory integration. ASR text is derived from audio, not an
independent sensory channel. TRIBE, MIRAGE and RR-stim receive no
measured BOLD. Internal neural projections, tokenizers and transitions remain
in the pixel-input controls.

\paragraph{Numerical precision.}
Eleven NeuroWorld input-control endpoints required FP32 fitting after
nonfinite-gradient failures under BF16: HCP V and VT at seed1; Algonauts V
at seed0 (sub-01/02/03), seed1 (sub-02/03/05), and seed2 (sub-02/03); and
Algonauts T at seed0 (sub-01). Seeds, learning rates, architectures, batches,
stages and validation criteria were unchanged; other endpoints retain BF16.
Recovery results were not selected by test scores. MLP/GRU learning-rate
selection uses finite completed candidates on the prescribed validation grid.

\subsection{Visual representations across models and datasets}
\label{app:visual_representation_all}
\paragraph{Pixel construction and ridge representation controls.}
\label{app:representation_controls}
All designated HCP, BOLD and Algonauts folds compare two inputs: AR plus 768 raw RGB values and AR plus 2,816 V-JEPA features. Each arm is jointly refitted with the same BOLD-history input and the same complete-coverage validation selector. Raw frames are resized to $16\times16$ at 2 Hz, with area interpolation, scaled by 255, then averaged in the same dataset-specific TR bins as the cached video representation. HCP reads the correct version's full movie and slices the already declared stimulus TRs; Algonauts reads source paths and offsets from the audited official event table. No additional HRF shift is applied. Pixel features use training-only normalization; the resulting BOLD normalizer must exactly match the source. All arms share the same target references, context, K, validation origins, test H, and ROI reduction. Audio and text are absent in all three arms. This is a representation contrast, not a dimension-matched attribution of the entire encoder effect to pretraining. Original pilot coefficients are preserved: HCP uses FP64 sufficient statistics and a symmetric eigensolve; BOLD fold0 and Algonauts sub-01 retain FP32 Cholesky fits. Newly completed BOLD folds 1--4 and Algonauts sub-02/sub-03/sub-05 use FP64 fitting and FP32 inference, with normalization recomputed from each fold's training subjects. The objective and validation-only penalty-selection rule are shared, but numerical solvers are not identical across all folds. HCP's pixel arm selects a penalty near $10^{-11}$ after prescribed boundary expansion. All arms cover the same 14 H, including BOLD's extended horizons. Results average subjects, not folds when their sizes differ; deterministic fits are not counted as three stochastic seed replications.

Figures~\ref{fig:matched_visual_all_app}
gives the full eight-model comparisons underlying Figure~\ref{fig:matched_visual_inputs}.
No model is omitted based on the size or direction of its encoder benefit.

\begin{figure}[htbp]
\centering
\includegraphics[width=\linewidth]{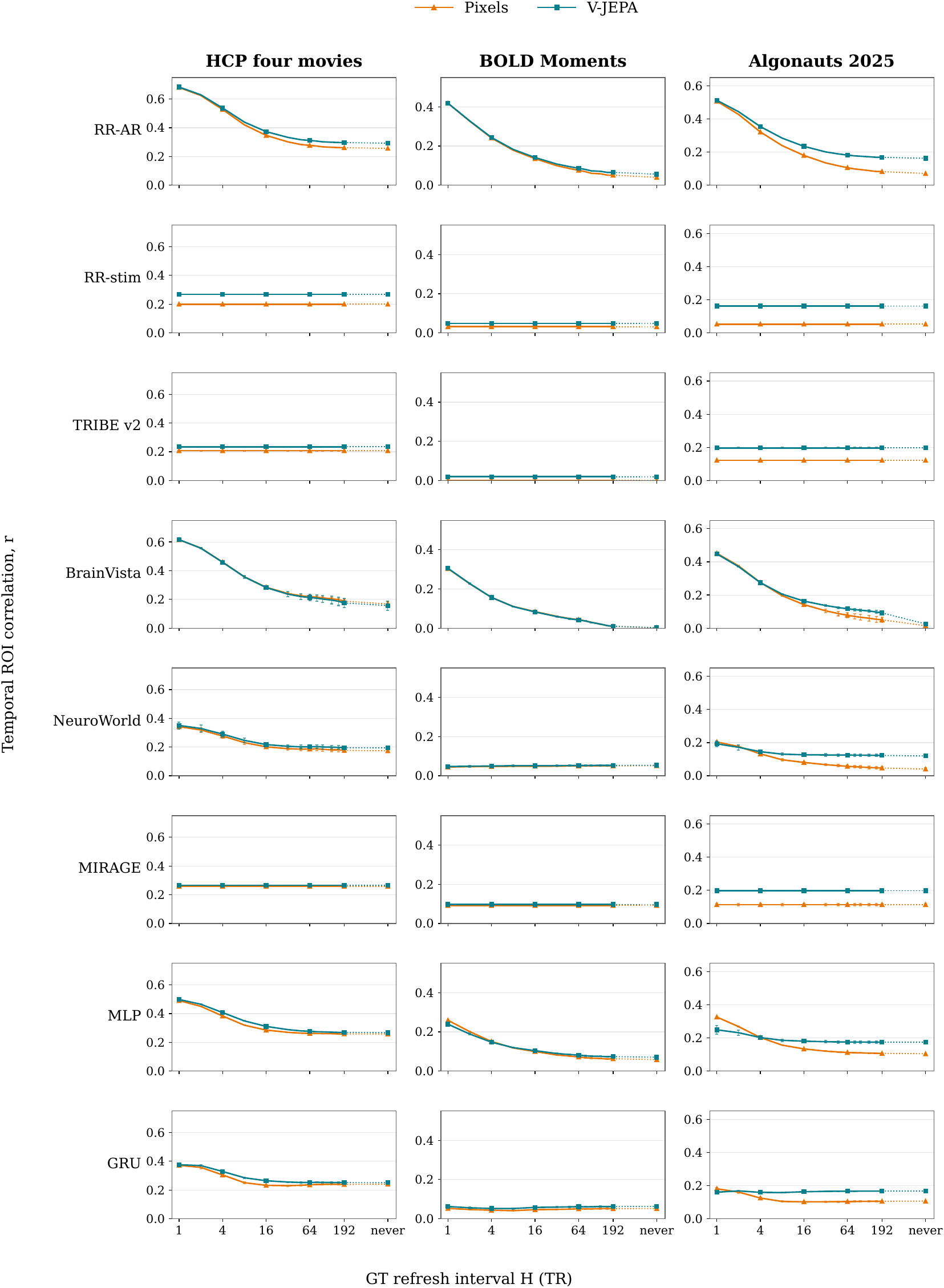}
\caption{Complete eight-model pixels versus V-JEPA comparison on HCP, BOLD Moments and Algonauts 2025 Datasets.
All designated held-out subjects and neural seeds 0/1/2 are included.
Bars show seed SD; RR-AR and RR-stim are deterministic. Dotted connectors
lead to the categorical no-refresh endpoint (\emph{never}).}
\label{fig:matched_visual_all_app}
\end{figure}

\subsection{Modality combinations}\label{app:modality}
The following figures compare the same stimulus inputs across predictors.
Full video/audio/text results are shown in Figure~\ref{fig:baseline};
Figures~\ref{fig:cross_model_single_modalities_app}
and~\ref{fig:cross_model_paired_modalities_app} show the single-
and paired-modality controls, respectively.

\begin{figure}[H]
\centering
\includegraphics[width=\linewidth]{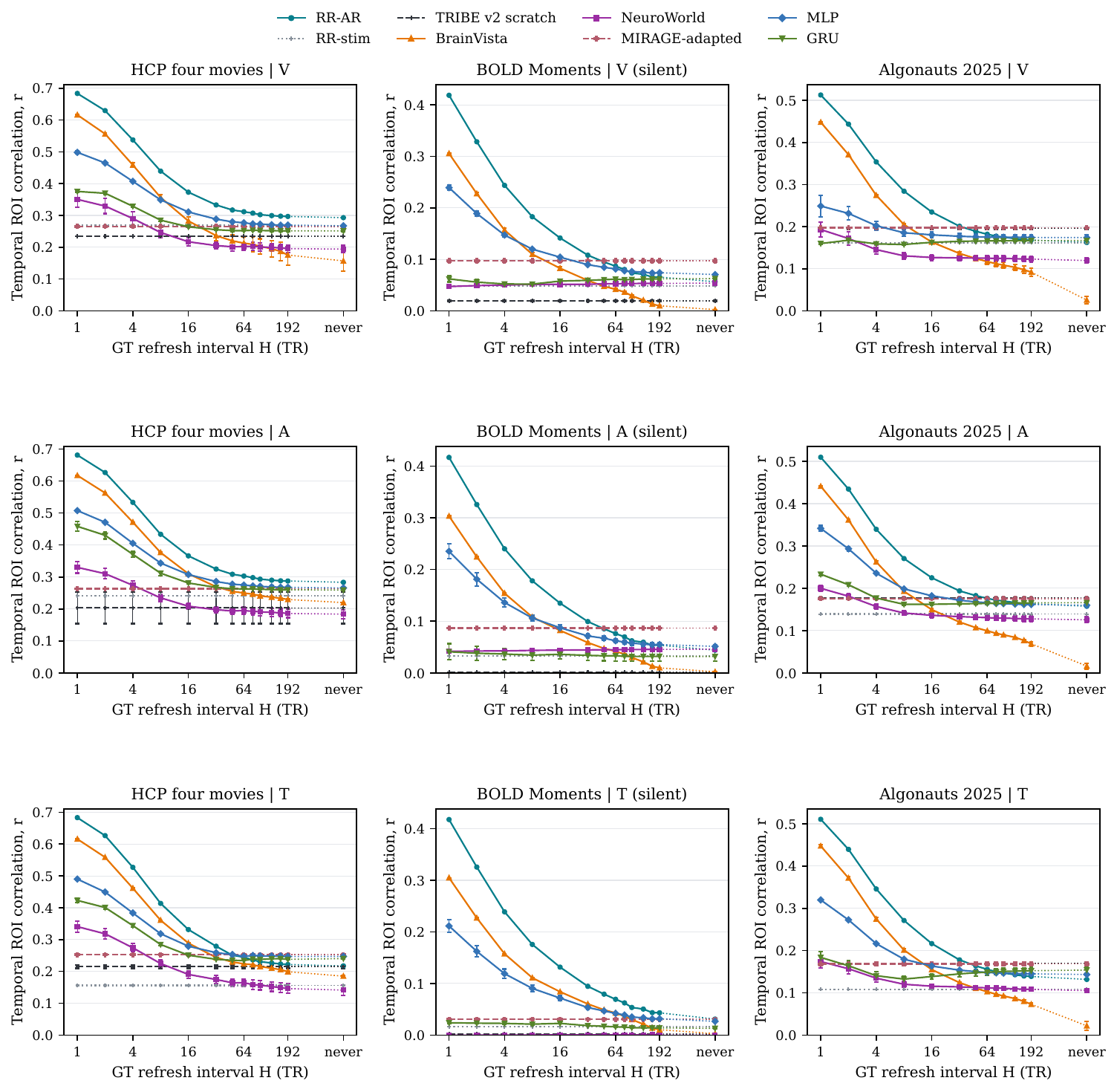}
\caption{Matched V, A and T inputs across datasets and models. Every displayed neural curve includes all designated held-out subjects and seeds 0/1/2; bars show sample SD of the three subject-equal seed means. Ridge fits are deterministic. All model--input combinations are complete; incomplete seeds or folds are not pooled. The y-axis is absolute temporal ROI correlation; finite H extends to 192, with dotted guides to the categorical never endpoint. BOLD was presented silently, so A/T are source-feature controls. TRIBE and RR-stim are history-free references.}
\label{fig:cross_model_single_modalities_app}
\end{figure}
\begin{figure}[H]
\centering
\includegraphics[width=\linewidth]{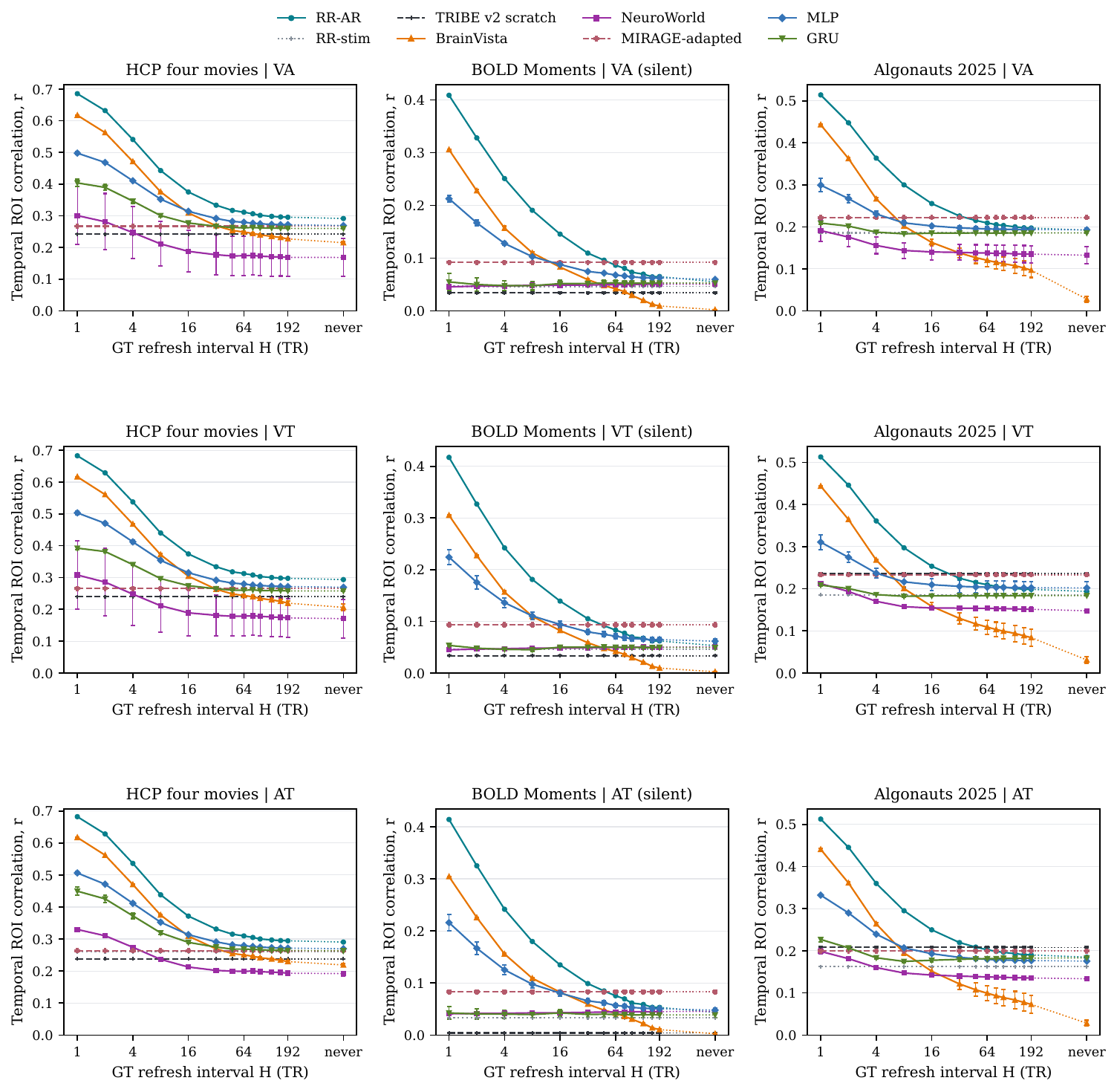}
\caption{Matched VA, VT and AT controls. All models cover every designated fold and seeds 0/1/2; bars are sample SD of subject-equal seed means. Ridge fits are deterministic. NeuroWorld includes the explicitly documented FP32 recovery fits. BOLD was presented silently.}
\label{fig:cross_model_paired_modalities_app}
\end{figure}

\subsection{Generalization across stimulus encoders}
\label{app:general_stimenc}

To test whether the benefit of pretrained stimulus representations is
specific to V-JEPA, we additionally evaluate DINOv2
\citep{oquab2023dinov2} and CLIP, using an OpenCLIP checkpoint
\citep{cherti2023reproducible}, on HCP four movies.
We compare these frozen visual representations with pixels using
RR-AR, RR-stim, and MLP, retaining the same subject split and evaluation
protocol.
Figure~\ref{fig:hcp_four} shows that all three pretrained encoders
improve prediction over pixels, with clearer gains when measured
history is sparse or unavailable.
In contrast, performance differences among the pretrained encoders
are comparatively small, particularly for RR-AR and MLP.
Thus, the benefit of richer stimulus representations is not confined
to V-JEPA: the main distinction here is between pixels and pretrained
features, rather than between individual pretrained encoders.

\begin{figure}[H]
\centering
\includegraphics[width=\linewidth]{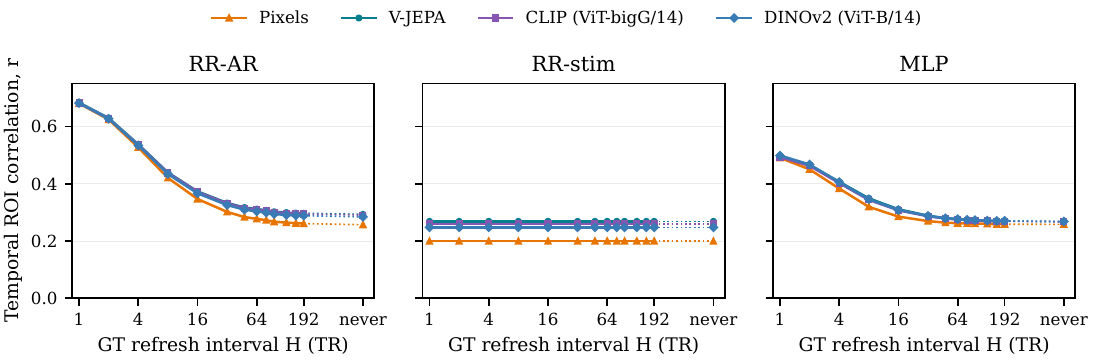}
\caption{
Generalization across visual encoders on HCP four movies.
Each panel compares pixels, V-JEPA, CLIP (ViT-bigG/14), and DINOv2
(ViT-B/14) within one predictor.
Curves show temporal ROI correlation across 32 held-out subjects
and 1,000 parcels.
MLP results are averaged over seeds 0/1/2, with bars indicating
sample SD across seed-level means; deterministic ridge fits have
no seed error bars.
The categorical \emph{never} endpoint denotes no further measured-history
refresh; dotted connectors are visual guides.
Feature dimensions and preprocessing are encoder-specific, so this
comparison evaluates representation pipelines rather than a
dimension-matched encoder ablation.
}
\label{fig:hcp_four}
\end{figure}

\section{Robustness to Context, Training and Model Capacity}
\label{app:diagnostics}
\label{app:window_controls}
\label{app:diagnostic_numbers}
\label{app:completed_followup}
These controls vary the comparison settings without changing the main
benchmark. C/K, regularization and solver controls use HCP fold0, BOLD fold0
and Algonauts sub-01; reduced-rank controls use all designated held-out subjects.

\subsection{Context length}
\label{sec:rr_context_diagnostic}
\label{sec:nn_window_diagnostic}

We compare $C\in\{1,4,8\}$ while holding the benchmark training interval fixed. Every dataset includes all eight models: deterministic RR-AR and RR-stim, together with the six neural models evaluated using seeds 0/1/2. RR-AR is refitted at each context length, whereas RR-stim is unchanged because it does not use measured BOLD history. The benchmark $C=8$ references are evaluated on the same held-out fold as the $C=1$ and $C=4$ controls rather than being pooled across additional test folds. Across $C\in\{1,4,8\}$, RR-AR remains competitive and generally outperforms the nonlinear history-conditioned predictors on these diagnostic splits. Stimulus-only MIRAGE and TRIBE v2 can match or exceed it at longer refresh intervals (Figure~\ref{fig:context_absolute}).

\begin{figure}[htbp]
\centering
\includegraphics[width=\linewidth]{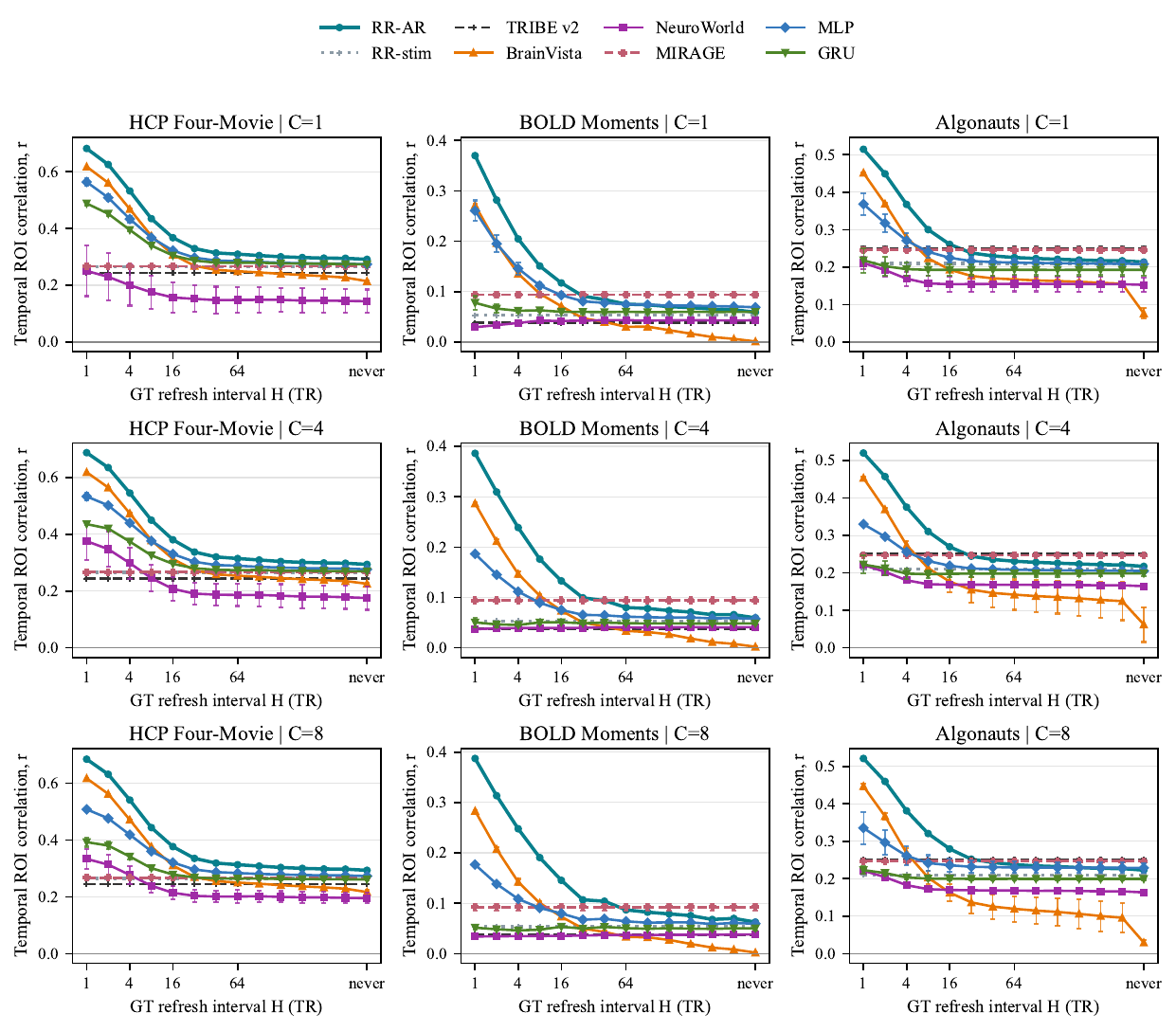}
\caption{{Context-length controls for $C\in\{1,4,8\}$ (rows), with the benchmark training interval fixed. Columns show HCP four movies, BOLD Moments, and Algonauts 2025 using HCP/BOLD fold0 and Algonauts sub-01. All six neural models use seeds 0/1/2; error bars denote sample SD across seeds. Deterministic ridge fits have no seed error bars. Benchmark references use the same held-out fold as the corresponding context-length controls. Both ridge references, all six neural models, and all 14 refresh settings are included. RR-stim does not use measured BOLD context and is therefore unchanged across rows.}}
\label{fig:context_absolute}
\end{figure}

\subsection{History-only controls}
\label{app:history_only_controls}
{{To isolate predictability from measured BOLD history, we refit RR-AR, BrainVista, MLP, and GRU after replacing all stimulus values with zero.  We compare each history-only fit with its corresponding main fit (history plus
stimulus) and with the three available history-free stimulus predictors
(RR-stim, TRIBE v2 and MIRAGE). The evaluation grid and reduction are unchanged
from the main benchmark. RR-AR is deterministic; the neural models use seeds
0/1/2. Results cover 32 HCP subjects, 10 BOLD Moments subjects and all four
designated Algonauts subjects.}}

{NeuroWorld is not included in the history-only ablation. Its two-stage procedure jointly learns a BOLD encoder and a latent predictor before freezing them to train a rollout decoder; under stimulus removal, representation degeneration can confound the effect of removing stimulus information. We therefore exclude this control from both real-data and simulation comparisons.}

{Figure~\ref{fig:history_stimulus_main} shows that the contribution of stimulus input varies across models, datasets, and refresh intervals. History-only RR-AR closely matches the full model at short intervals across all three datasets, but its accuracy declines as the interval increases, producing a clear full-model advantage at long intervals on HCP and Algonauts. BrainVista shows a clearer long-interval benefit from stimulus input on HCP than on BOLD Moments or Algonauts, where the full and history-only curves remain closer. For MLP and GRU, history-only fits can match or exceed the full model at some short intervals on BOLD Moments and Algonauts, while stimulus input generally becomes more useful at longer intervals. The stimulus-only references remain unchanged across refresh intervals and can overtake history-only predictors as measured-history refreshes become less frequent. Together, these results show that recent BOLD history supports much of the short-interval accuracy, while the incremental benefit of stimulus input depends on both the model and the dataset.}

\begin{figure}[htbp]
\centering
\includegraphics[width=\linewidth]{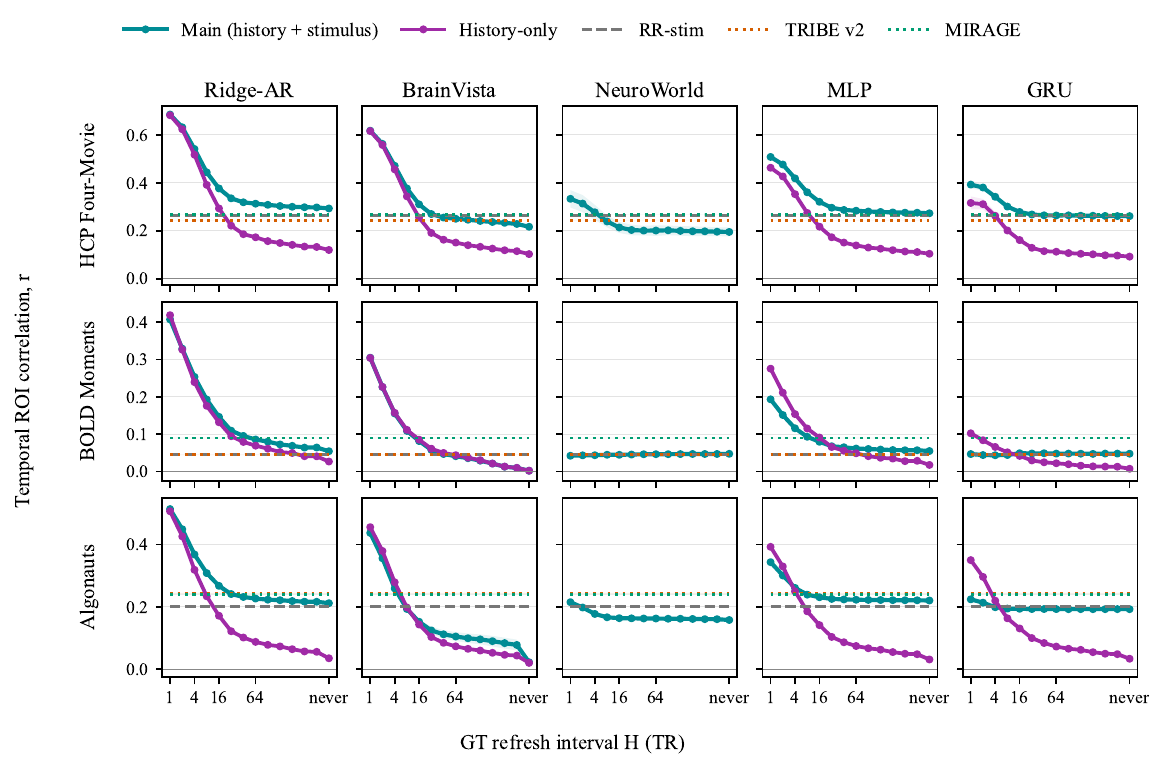}
\caption{{{History-only, main and stimulus-only comparisons. Each panel shows that column's main model (history plus stimulus), with history-only ablations for RR-AR, BrainVista, MLP, and GRU, alongside the same three available stimulus-only references: RR-stim, TRIBE v2 and MIRAGE. Neural curves show the mean over seeds 0/1/2 with sample-SD shading; ridge fits are deterministic. BOLD Moments aggregates all five held-out folds, and Algonauts aggregates sub-01, sub-02, sub-03 and sub-05. NeuroWorld is shown with its main inputs only; its history-only control is excluded for the methodological reason described above.}}}
\label{fig:history_stimulus_main}
\end{figure}

\subsection{{{Neural-model training interval}}}
\label{sec:rr_k_diagnostic}
{In Figure}~{\ref{fig:training_k_absolute}, we compare
\(K\in\{8,32\}\) with benchmark \(K=64\) for HCP/BOLD Moments or \(K=92\)
for Algonauts, retaining \(C=8\). The six neural models are trained and
selected separately at each \(K\), using seeds 0/1/2 on the same fixed folds
as the context controls. Training and checkpoint-selection \(K\) change
together; test refresh intervals and scored targets remain fixed. For TRIBE
v2 and MIRAGE, \(K\) specifies the supplied stimulus/target interval rather
than recurrent BPTT depth.}

RR-AR and RR-stim retain their benchmark coefficients and selected penalties throughout these neural-model controls. Selecting ridge penalties using a different validation interval could change the final ridge predictor, even though the candidate one-step fits are independent of $K$. Changing the neural training interval affects accuracy and some model rankings, but does not produce a consistent advantage over the fixed RR-AR benchmark on these diagnostic splits (Figure~\ref{fig:training_k_absolute}). BrainVista is among the strongest neural predictors at short refresh intervals, although its accuracy generally declines faster than RR-AR's as $H$ increases. At longer intervals, MIRAGE can overtake RR-AR on BOLD Moments, as can MIRAGE and TRIBE v2 on Algonauts.

\begin{figure}[htbp]
\centering
\includegraphics[width=\linewidth]{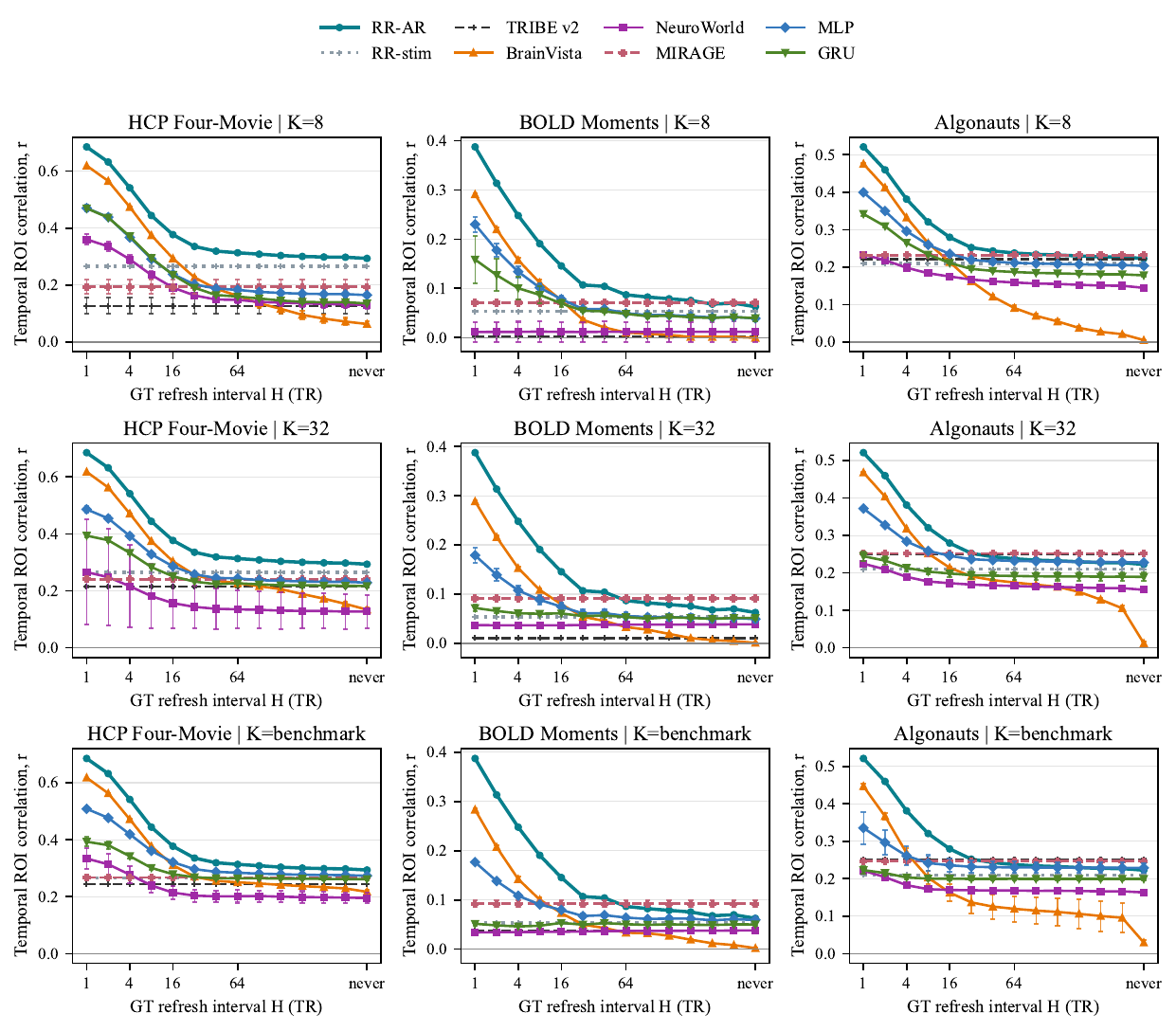}
\caption{{{Neural-model training-interval controls for \(K=8/32/\)benchmark (rows), with \(C=8\) throughout. Benchmark \(K\) is 64 for HCP/BOLD Moments and 92 for Algonauts. The six neural models are trained separately at each \(K\) using seeds 0/1/2. RR-AR and RR-stim are fixed benchmark references repeated across rows.}}}
\label{fig:training_k_absolute}
\end{figure}

\subsection{Optimization controls: mini-batch versus closed-form fitting}
\label{sec:rr_fitting_diagnostic}
Could RR's predictive advantage depend on solving its coefficients directly rather than learning them with mini-batches? We fix the linear model, all one-step training rows, C=8, input features, training-only normalization, splits, and the originally validation-selected penalty, then train the coefficients and unpenalized intercept from zero with Adam. No fitted RR weights or teacher predictions are used in the updates. The experiment covers HCP fold0, BOLD fold0 and Algonauts sub-01, each with seeds 0/1/2.

\paragraph{Matched objective and training protocol.}
For design matrix $Z$, targets $Y$, $N$ training rows and $R$ ROIs, both solvers minimize
\begin{equation}
J(W,b)=\frac{\|ZW+\mathbf{1}b^\top-Y\|_F^2}{NR}
       +\frac{\lambda}{R}\|W\|_F^2.
\label{eq:minibatch_control}
\end{equation}
This is the original one-step ridge objective scaled by $1/R$, not a K-step BPTT loss. Fixed penalties are 0.1, 10 and 1 for HCP, BOLD and Algonauts, respectively. Each epoch shuffles and visits every training row once in batches of 1,024, retaining the final short batch. Adam uses learning rate $10^{-3}$, $\beta=(0.9,0.999)$, $\epsilon=10^{-8}$ and zero optimizer weight decay; the L2 term is explicit in Eq.~\ref{eq:minibatch_control}. A ReduceLROnPlateau schedule monitors only the exact full-training objective (factor 0.2, patience 8, relative improvement threshold $10^{-5}$, minimum learning rate $10^{-7}$). Training uses FP32 without AMP or TF32. A fresh FP64 Cholesky fit supplies the optimal objective $J^*$ for numerical audits, never optimization targets or initial parameters. Both methods use FP32 inference and the original validation/test evaluator.

The prespecified stopping certificate requires both $(J-J^*)/(J_0-J^*)\le10^{-6}$ and a relative normal-equation residual, including the bias equation, at most $10^{-4}$; $J_0$ is the zero-initialized objective. Otherwise training stops at 300 epochs. The terminal iterate is evaluated at every standard and extended H, without validation or test checkpoint selection. The full-training objective and convergence audit are not computed from held-out targets.

\paragraph{Result and conclusion boundary.}
All nine runs reach the 300-epoch ceiling without satisfying both numerical criteria. Nevertheless, their predictive performance nearly coincides with the closed-form reference (Figure~\ref{fig:solver_absolute}). {The maximum absolute correlation discrepancy over all three seeds and all 14 H conditions is $3.44\times10^{-5}$ on HCP, $1.96\times10^{-4}$ on BOLD, and $2.51\times10^{-5}$ on Algonauts.} The largest corresponding scalar-mean regularized objective gaps are $2.40\times10^{-6}$, $3.85\times10^{-5}$ and $1.10\times10^{-6}$. Fresh closed-form correlations agree with the original RR at standard H to within $5\times10^{-8}$. Thus mini-batch optimization from zero reproduces the linear reference's predictive performance in these tested folds: closed-form solving is not necessary for that performance. This is not certified numerical convergence, a formal equivalence test, or evidence that nonlinear optimization is adequate.

\begin{figure}[htbp]
\centering
\includegraphics[width=\linewidth]{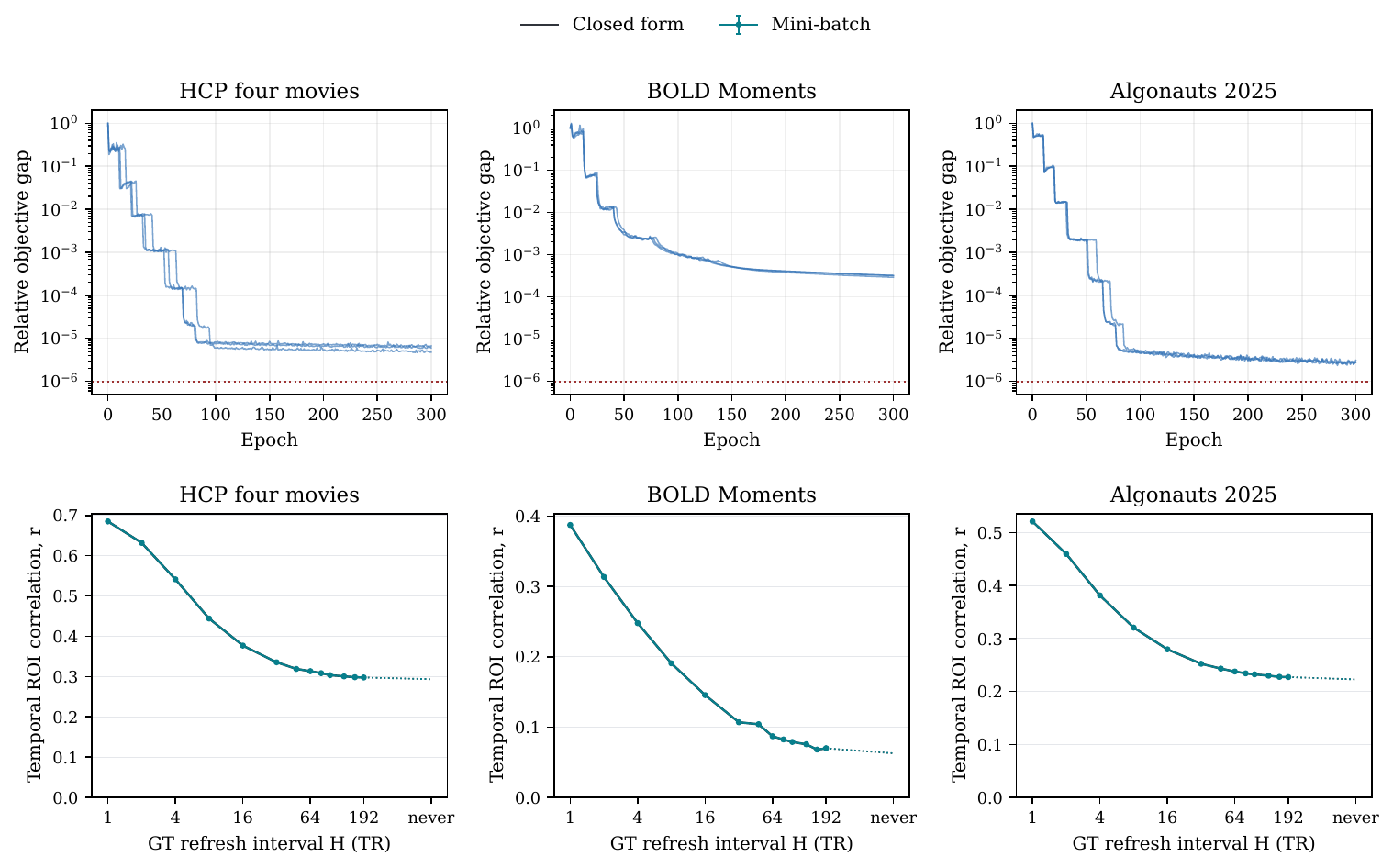}
\caption{Same-objective mini-batch optimization control, HCP/BOLD fold0 and Algonauts sub-01, seeds 0/1/2. Top: $(J-J^*)/(J_0-J^*)$ during zero-initialized Adam training; thin lines are individual seeds and the thick line is their mean. The dotted line is the prespecified objective-gap threshold, not a performance-equivalence margin. All nine runs reach 300 epochs without satisfying both numerical convergence criteria. Bottom: absolute temporal ROI correlation of the terminal mini-batch models (mean $\pm$ sample SD across three seeds) and the fresh closed-form reference. The near-overlap reflects predictive agreement, not certified convergence. Dotted connectors to never are visual guides to a categorical finite-record condition. No validation/test checkpoint selection or RR initialization is used.}
\label{fig:solver_absolute}
\end{figure}

\subsection{Regularization strength}
\label{app:regularization_curves}
Figure~\ref{fig:regularization} displays the results across different regularization strength. The fixed diagnostic splits are HCP fold0, BOLD fold0 and Algonauts sub-01. We evaluate every prespecified $\lambda\in\{0,\allowbreak 0.01,\allowbreak 0.1,\allowbreak 1,\allowbreak 10,\allowbreak 100\}$ on the same design. All endpoints are reported at the standard H values and at 128,160,192 and never. The results show that the performance is sensitive to the regularization strength.

\begin{figure}[htbp]
\centering
\includegraphics[width=\linewidth]{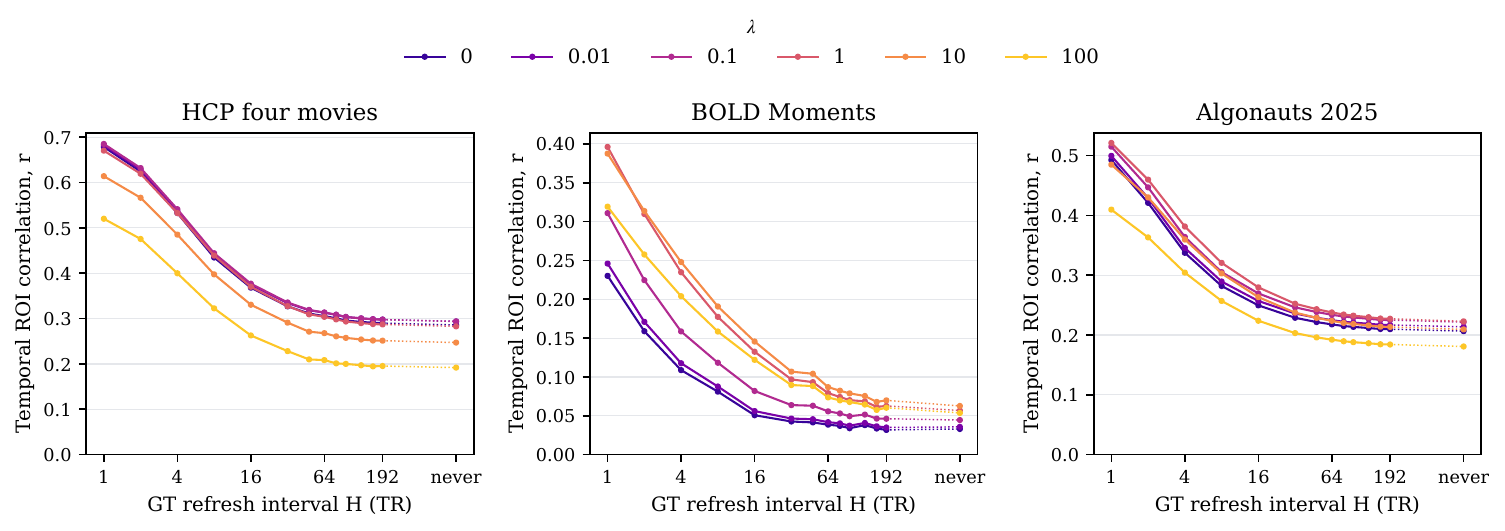}
\caption{Fixed-grid regularization controls, seed0, one declared fold per dataset. Every tested penalty is shown through H=192 and never, the finite-record no-refresh endpoint.}
\label{fig:regularization}
\end{figure}

\subsection{Reduced-rank controls}
\label{app:reduced_rank_controls}

All fixed ranks are reported. Coefficient counts include the intercept but exclude frozen stimulus encoders; they count stored factors, not intrinsic degrees of freedom. These are deterministic seed0 fits.

{\color{black}
The main benchmark uses dense RR-AR. Here we constrain the rank of its joint history--stimulus coefficient matrix under the same ridge objective. Figure~\ref{fig:lowrank_absolute} shows that substantial compression preserves much of its long-horizon accuracy. The estimator is described below, and parameter accounting is detailed in Appendix~\ref{app:parameter_counts}.
}

\paragraph{Estimator and implementation.}
\label{app:lowrank_estimation}
We minimize the objective in Eq.~\ref{eq:ridge_fit} subject to \(\operatorname{rank}(W)\le r\). If \(V_r\) contains the leading orthonormal eigenvectors of \(Q^\top W_\lambda\), the reduced-rank solution is
\begin{equation}
W_{\lambda,r}=(W_\lambda V_r)V_r^\top.
\label{eq:lowrank_fit}
\end{equation}
Completing the square in the ridge objective gives a rank-\(r\) approximation in the \((G+\lambda I)\) metric, rather than ordinary singular-value truncation of the coefficient matrix. The zero-penalty case uses the numerical minimum-norm convention. The factorized map stores \((d+R)r+R\) coefficients including the unpenalized intercept, versus \(dR+R\) for the dense map. Regularization is selected separately on validation data for each prespecified rank. Fitting uses FP64 and prediction FP32, with all designated folds included. Both ridge regression and its reduced-rank form are established estimators; we use them as controls on predictive model complexity.

\begin{figure}[htbp]
\centering
\includegraphics[width=\linewidth]{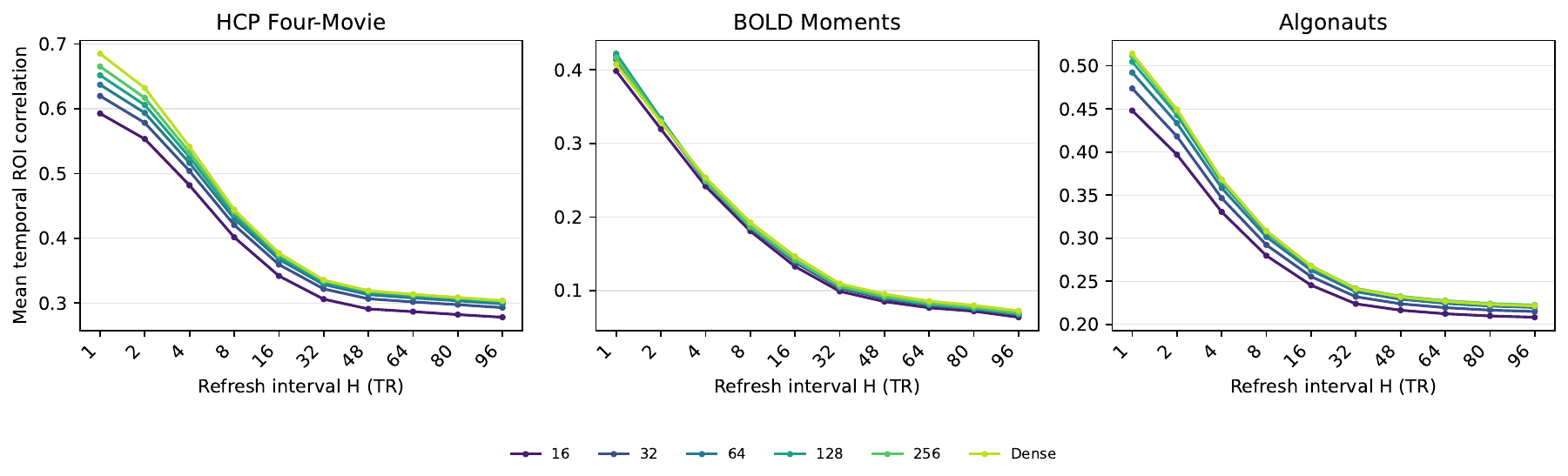}
\caption{{\color{black} Reduced-rank controls for the joint RR-AR predictor. All prespecified ranks and the matched dense fit use video, audio, and text features, with C=8 and separately validation-selected ridge penalties. Curves show absolute temporal ROI correlation over all designated folds at H=1--96. Fits are deterministic; seed0 does not denote independent replication. Parameter accounting follows Appendix~\ref{app:parameter_counts}.}}
\label{fig:lowrank_absolute}
\end{figure}

\section{Regional Predictability and Cross-Subject Analyses}
\label{app:regional_section}
\label{app:regional}

This section supports the main-text regional analysis with two complementary
questions: whether predicted trajectories preserve large-scale functional
organization, and how regional prediction accuracy relates to response
consistency across training subjects. Unless stated otherwise, regional analyses include every designated fold, all 1,000 cortical targets in each dataset, and seeds 0/1/2.

For each parcel, we first average its temporal prediction accuracy across valid
clips from the same held-out subject and then average equally across held-out
subjects. When comparing two models, we use exactly the same subjects, records, parcels,
prediction horizons, and valid time points. All 1,000 parcels are included. 

\subsection{Preservation of functional organization}
\label{app:regional_fc}

Temporal correlation evaluates each parcel separately. We additionally ask
whether the joint predicted trajectory preserves the correlation structure
among parcels. For each continuous held-out record and prediction at $H=32$,
we compute the parcel-by-parcel Pearson functional-connectivity (FC) matrix for
both the prediction and its matched ground truth. Off-diagonal correlations are
averaged in Fisher-$z$ space within subject. 

For the mean-FC display in Figure~\ref{fig:fc_h32_all}, these matrices are averaged equally over subjects and, where applicable, seeds before applying the inverse Fisher transform. In the HCP row, lower triangles show group-average FC in Pearson-$r$ units, whereas upper triangles show intersubject FC variability. For each edge, variability is computed as the sample standard deviation across subjects of the subject-level Fisher-$z$ FC values, after averaging across seeds within subject for stochastic models. Deterministic ridge models use their single fit. We report this variability only for HCP; BOLD Moments and Algonauts have smaller subject samples. Parcels are ordered by the seven Yeo networks.

The scalar FC similarity in Table~\ref{tab:fc_h32} is computed separately for every subject and seed as the Pearson correlation between the off-diagonal upper-triangular entries of the predicted and ground-truth FC matrices; seed-specific values are averaged within subject, and the table reports the mean and sample SD across subjects. This subject-first statistic is intentionally not the correlation between two group-average matrices.

\begin{figure}[htbp]
\centering
\includegraphics[width=\linewidth]{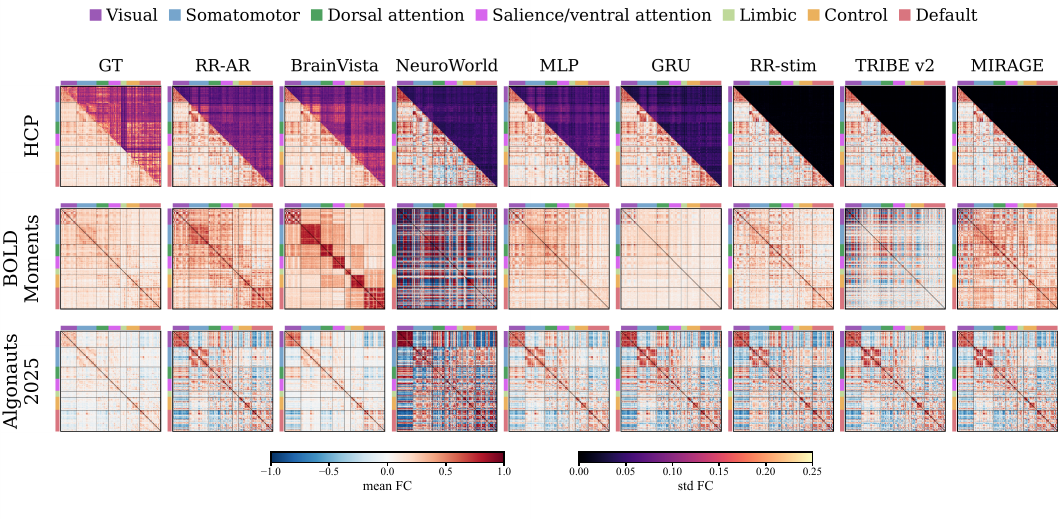}
\caption{Ground-truth and predicted functional connectivity at $H=32$ across three datasets, with parcels ordered by Yeo-7 network. In the HCP row, lower triangles show group-average FC in Pearson-$r$ units (left colorbar), and upper triangles show intersubject variability, quantified as the sample SD across subjects of subject-level Fisher-$z$ FC values after averaging across seeds within subject (right colorbar, range 0--0.25).
BOLD Moments and Algonauts show mean FC in both triangles;
intersubject variability is reported only for HCP.}
\label{fig:fc_h32_all}
\end{figure}

Figure~\ref{fig:fc_h32_all} shows that the predictions recover broad within- and between-network organization rather than merely matching each parcel's marginal time course. The amount of recovered structure nevertheless varies across models and datasets. RR-AR is among the strongest FC predictors in all three datasets, with subject-level similarities of $0.730\pm0.042$, $0.626\pm0.045$, and $0.691\pm0.008$ for HCP, BOLD Moments, and Algonauts, respectively (Table~\ref{tab:fc_h32}). Relative to its stimulus-only counterpart, RR-AR improves FC similarity by 0.142 on HCP, 0.188 on BOLD Moments, and 0.094 on Algonauts. The advantage is not exclusive: BrainVista is close to RR-AR on HCP and Algonauts, and MLP is close on Algonauts. Conversely, methods with similar temporal-accuracy curves can differ visibly in their recovered FC patterns. On HCP, history-conditioned predictors produce appreciable intersubject FC variability, whereas
stimulus-only predictors show little to no variability across subjects receiving the same stimulus inputs (Figure~\ref{fig:fc_h32_all}, upper triangles). FC preservation therefore supplies a complementary systems-level criterion, rather than a restatement of parcelwise prediction accuracy or evidence that the learned predictor identifies causal connectivity.

\begin{table}[htbp]
\centering
\small
\caption{Similarity between predicted and ground-truth functional
connectivity at $H=32$, measured by Pearson correlation between their
off-diagonal upper-triangular entries. Values report the mean $\pm$
sample SD across held-out subjects after averaging the three
seed-specific correlations within each subject.}
\label{tab:fc_h32}
\setlength{\tabcolsep}{4pt}
\renewcommand{\arraystretch}{1.15}
\resizebox{\linewidth}{!}{%
\begin{tabular}{@{}lcccccccc@{}}
\toprule
Dataset
& RR-AR
& RR-stim
& \shortstack{TRIBE v2}
& BrainVista
& NeuroWorld
& \shortstack{MIRAGE}
& MLP
& GRU \\
\midrule
HCP four movies
& $\mathbf{0.73 \pm 0.04}$
& $0.59 \pm 0.05$
& $0.56 \pm 0.05$
& $0.72 \pm 0.03$
& $0.48 \pm 0.04$
& $0.58 \pm 0.05$
& $0.65 \pm 0.04$
& $0.63 \pm 0.05$ \\

BOLD Moments
& $\mathbf{0.63 \pm 0.05}$
& $0.44 \pm 0.04$
& $0.18 \pm 0.03$
& $0.47 \pm 0.04$
& $0.19 \pm 0.03$
& $0.39 \pm 0.03$
& $0.38 \pm 0.03$
& $0.35 \pm 0.05$ \\

Algonauts 2025
& $0.69 \pm 0.01$
& $0.60 \pm 0.01$
& $0.63 \pm 0.01$
& $\mathbf{0.70 \pm 0.01}$
& $0.47 \pm 0.01$
& $0.64 \pm 0.01$
& $0.69 \pm 0.01$
& $0.64 \pm 0.01$ \\
\bottomrule
\end{tabular}%
}
\end{table}
\subsection{Training-subject response consistency and held-out regional prediction}
\label{app:regional_isc}

Figure~\ref{fig:regional_isc_h32} in Section~\ref{sec:shared_results} relates
training-subject response consistency to held-out parcel accuracy and to the
gain from measured history at $H=32$. Here we specify the alignment and
aggregation, and use $H=96$ as a longer-horizon sensitivity analysis.

\paragraph{Alignment without held-out leakage.}
For each evaluation fold, response consistency is computed only from its
training subjects. HCP uses 36 aligned timeline groups from 107 training
subjects and evaluates 32 held-out subjects. Each of five BOLD Moments folds
uses eight training subjects and evaluates two held-out subjects. Each of four
Algonauts LOSO folds uses three training subjects across 297 aligned groups and
evaluates the remaining subject. HCP and Algonauts groups match stimulus
identity and time coordinate.

BOLD Moments requires event rather than run alignment because presentation
order varies across subjects. Each event is sampled at onset plus 1.5 seconds
on the response grid that has already been shifted by three TRs. Linear
interpolation uses the two neighboring samples only when both lie within the
event and valid analysis interval; no additional HRF shift is applied.
Repeated presentations are averaged within training subject, after which
shared video identities are compared in a common order. The BOLD Moments
quantity is therefore event-pattern correspondence, not temporal ISC.

\paragraph{Estimation and aggregation.}
Within every aligned group and parcel, each training subject's response is
centered and standardized over matched samples and correlated with the mean
standardized response of the remaining training subjects. Valid correlations
are averaged in Fisher-$z$ space over subjects and then equally over groups,
followed by the inverse transform. Constant responses and constant peer means
are omitted. Each held-out subject's parcel accuracy is paired with the
response-consistency estimate from that subject's training fold. Parcel values
are averaged equally over held-out subjects, and descriptive Spearman
associations are computed across all 1,000 parcels.

\begin{table}[htbp]
\centering\small
\caption{Across-parcel Spearman associations between training-subject response
consistency and held-out prediction at $H=96$. History gain is RR-AR minus
RR-stim. These longer-horizon summaries are descriptive spatial associations,
not tests over independent parcels.}
\label{tab:regional_isc_association}
\begin{tabular}{lrrrrr}
\toprule
Dataset & Subjects & Parcels & RR-AR & RR-stim & History gain\\
\midrule
HCP four movies & 32 & 1,000 & 0.986 & 0.989 & $-0.498$\\
BOLD Moments & 10 & 1,000 & 0.809 & 0.871 & $-0.381$\\
Algonauts 2025 & 4 & 1,000 & 0.969 & 0.962 & $-0.170$\\
\bottomrule
\end{tabular}
\end{table}

The main-text $H=32$ result and the $H=96$ sensitivity analysis give the same
two-part picture. First, parcels with more consistent training-subject
responses are easier to predict with both RR-AR and RR-stim; the relationship
therefore characterizes general regional predictability rather than uniquely
explaining the linear--neural model gap. Second, measured history contributes a
different spatial profile. At $H=96$, consistency is negatively associated
with RR-AR minus RR-stim in every dataset, most strongly on HCP and least
strongly on Algonauts. The association is not a universal inverse law: it
varies across subjects and within networks, and Control or Default need not
show the largest gain in every dataset.

Response consistency measures correspondence, not activation magnitude, raw
intersubject variance, or temporal SNR. Prepared-signal variance is not
comparable across datasets, particularly because Algonauts applies upstream
whole-run standardization. Together, the main-text and longer-horizon analyses
support a limited conclusion: shared responses are generally easier to predict,
whereas measured history contributes a complementary regional profile. They do
not identify a biological mechanism or uniquely explain the performance gap
between linear and neural predictors.

\section{Simulation Design for Separating Neural and Hemodynamic Time Scales}
\label{app:hrf_model_controls}

Empirical BOLD prediction combines two distinct temporal processes: the
evolution of the underlying neural state and its transformation through the
hemodynamic observation process. To separate these factors, we construct a
paired simulation in which identical neural trajectories are observed through
hemodynamic models with different time scales. The purpose of this simulation
is to test sensitivity to the observation process under controlled conditions,
rather than to reproduce the full biological complexity of measured fMRI.

\paragraph{Neural and stimulus dynamics.}
The simulator contains 11 audiovisual ROIs with fixed linear neural dynamics
driven by two stimulus channels representing visual and auditory input. Each
five-minute sequence begins with 20 seconds of fixation, followed by one of
three stimulus schedules: sustained stimulation for 280 seconds,
5 seconds on/25 seconds off, or 20 seconds on/20 seconds off. Fifty simulated
subjects are divided into fixed training, validation, and test sets of
35, 5, and 10 subjects, respectively.

\paragraph{Paired hemodynamic intervention.}
For every subject and stimulus schedule, the same latent neural trajectory is passed through fast, medium, and slow Balloon--Windkessel observation models, with nominal HRF times to peak of 2, 6, and 10 seconds, respectively. Within each condition, all ROIs share a common nominal HRF. The comparison therefore holds the stimuli, latent neural trajectories, and subject assignments fixed while varying the hemodynamic observation time scale. 
The neural state before hemodynamic transformation is sampled on the same 1-s grid and retained as a matched no-HRF reference. This reference is a simulated latent state; it should not be interpreted as an electrophysiological measurement, a deconvolved BOLD signal, or an estimate of latent neural activity from the empirical datasets.

\paragraph{Neural simulation, BOLD sampling, and refresh intervals.}
Neural dynamics are simulated with a time step of $\Delta t=5$~ms. The resulting neural trajectories drive the Balloon--Windkessel observation model, and its BOLD output is sampled at $\mathrm{TR}=1$~s for model fitting and evaluation. The matched no-HRF reference is sampled from the neural trajectory on the same 1-s grid. Thus, 5~ms specifies the neural simulation time step, while prediction intervals are defined on the 1-s sampling grid. RR-AR receives the two-dimensional synthetic stimulus drive directly, without a pretrained media encoder, and an observed context of $C=8$ samples (8~s). Refresh intervals $H\in\{1,2,4,8,16,32,48,64,80,96,128,192\}$ therefore correspond to 1--192~s; in particular, $H=32$ corresponds to 32~s. A rollout length of $K=64$ samples corresponds to 64~s. The train/validation/test split and scoring times are matched across observation conditions. RR-AR fitting is deterministic.

\paragraph{Interpretation and limitations.}
This paired design isolates how the observation time scale changes apparent
predictive persistence while holding the simulated neural trajectory fixed.
It does not establish that hemodynamics explains the empirical performance gap
between linear and neural predictors. In particular, the latent dynamics in
this intervention are linear, the system contains only 11 ROIs, and the
specified HRFs span a controlled family rather than the full variability of
human hemodynamics. Conclusions from this simulation should therefore be
interpreted as a sensitivity analysis of the measurement process, not as a
mechanistic model of the empirical datasets.

\begingroup
\section{Complete Model Comparisons on Simulated Neural and BOLD Data}
\label{app:hrf_all_models}

\noindent\textbf{Models and matched controls.}
{Figure}{\ref{fig:hrf_all_models_appendix} extends the RR-AR analysis in Section}~{\ref{sec:hrf_results} to all eight predictors under the simulation design in Appendix}~{\ref{app:hrf_model_controls}. We evaluate RR-AR, RR-stim, MLP, GRU, BrainVista, NeuroWorld, MIRAGE, and TRIBE v2 on the three stimulus schedules and four observation conditions. Predictors use the two-channel synthetic stimulus directly, without pretrained media encoders; history-conditioned models use $C=8$, with neural rollout training and validation at $K=64$. MIRAGE and TRIBE v2 use fixed $K=64$ evaluation chunks. We keep these windows fixed when varying $H$ so that the stimulus context supplied to these history-free models does not change. Neural predictors use seeds 0/1/2 and ridge fits are deterministic. History-only controls are refitted for RR-AR, MLP, GRU, and BrainVista with stimulus inputs zeroed throughout training, validation, and testing, retaining the same targets, splits, and normalization. We apply the same methodological exclusion of NeuroWorld history-only controls as in Appendix}~{\ref{app:history_only_controls}, retaining its full-input results. RR-stim, MIRAGE, and TRIBE v2 have no measured-history input in these adaptations and therefore have no corresponding history-only control.}

\medskip
\noindent\textbf{{Validation-selected training settings.}}
{We selected one configuration per neural model and input condition by mean validation $K=64$ MSE over six seed-0 pilot conditions: all four observations under sustained stimulation and slow-HRF observations under the two ON/OFF schedules. This 144-fit sweep did not use test targets; selected configurations were fixed for all 12 conditions and seeds 0/1/2, yielding 360 formal evaluations, including 60 reused pilot fits. The 36 NeuroWorld history-only evaluations are excluded from the reported comparison for consistency with the real-data ablation scope. MLP and GRU use learning rate $10^{-3}$ and at most 400 epochs. BrainVista uses tokenizer/predictor learning rates $(3\times10^{-4},10^{-4})$ and epoch limits $(1000,120)$. NeuroWorld uses latent/decoder rates $(2\times10^{-4},10^{-3})$ for full inputs, with limits $(150,300)$. MIRAGE and TRIBE v2 are trained from scratch at $10^{-4}$ for at most 60 epochs. All use patience 30 and the best validation checkpoint; ridge results are retained from the original fits. These are finite-budget selections, not guarantees of convergence: several tokenizer and stimulus-only fits attain their best validation loss at the epoch limit.}

\medskip
\noindent\textbf{{Dependence on refresh interval and HRF time scale.}}
{RR-AR remains competitive across the simulated conditions, with the highest mean correlation in all 12 conditions at $H=1$ and $H=4$, and in nine at $H=16$; several neural predictors exceed it at longer intervals (Figure}~{\ref{fig:hrf_all_models_appendix}). Slower HRFs also sustain stronger history-only predictability: at $H=4$ and $H=16$, history-only correlation increases from fast to medium to slow HRFs for RR-AR, BrainVista, MLP, and GRU under every stimulus schedule. This supports greater persistence of history-based predictability with slower hemodynamics, although the ordering does not hold uniformly at long intervals. Comparing each full model with its own history-only refit further shows that stimulus input becomes more valuable as measured-history refreshes become sparse: across RR-AR, BrainVista, MLP, and GRU and the 12 conditions, the mean full-minus-history-only correlation gap increases from $0.050$ at $H=1$ to $0.393$ without further refreshes, where all 48 paired comparisons favor full inputs. Thus history supports short-interval prediction, while stimulus input provides a larger incremental benefit over long intervals.}

\begin{figure}[htbp]
\centering
\includegraphics[width=\linewidth]{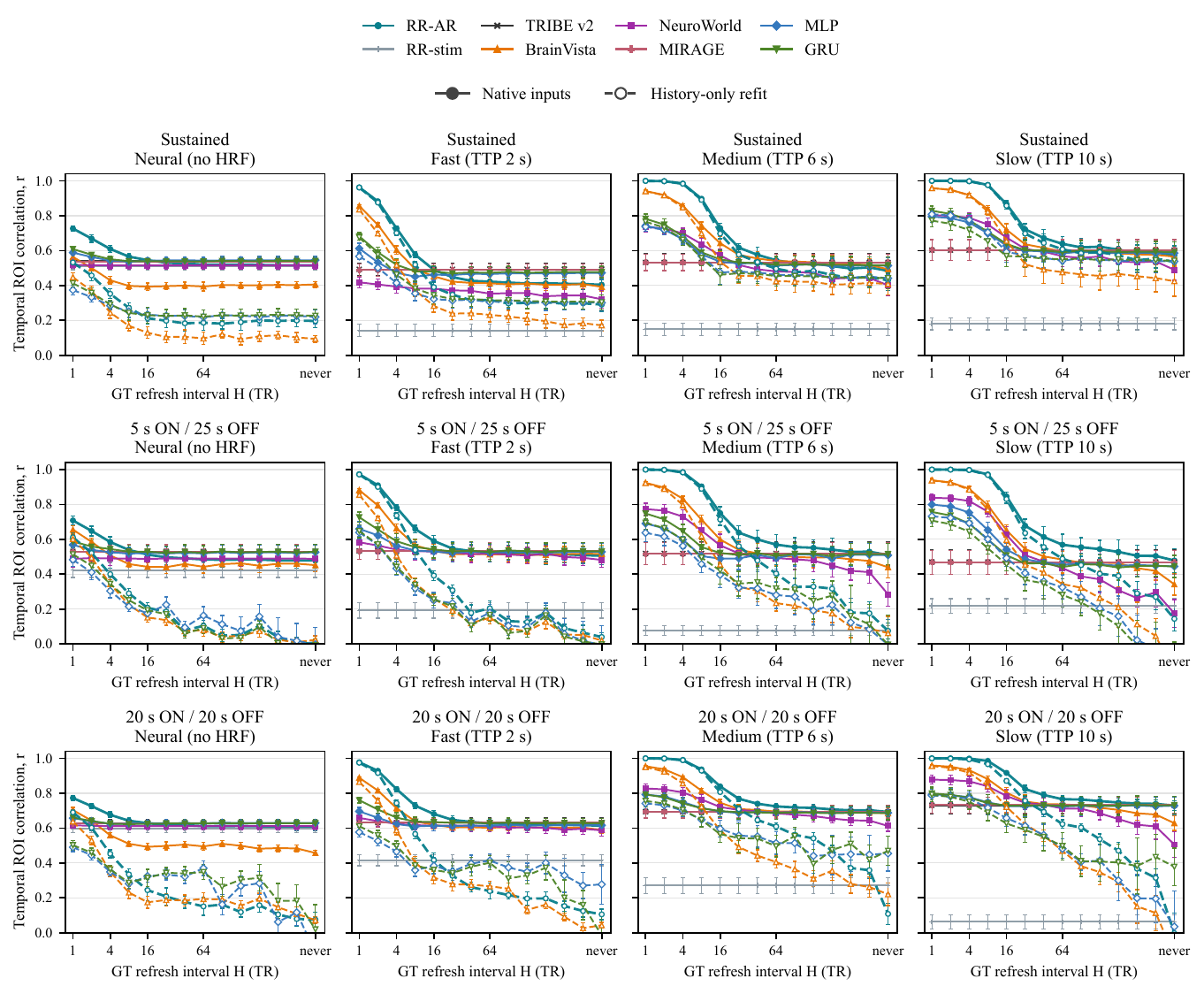}

\caption{Comparison of eight models and four history-only controls on simulated neural and BOLD data. Rows show sustained, 5 s ON/25 s OFF, and 20 s ON/20 s OFF stimulation; columns show the no-HRF neural reference and fast, medium, and slow HRFs. Solid lines show full-input predictions (stimulus only or history plus stimulus); dashed lines with open markers show history-only refits. Correlations are averaged over 11 ROIs and then across seeds within subject for neural models. Curves show means across 10 test subjects; error bars show SD across subjects. \emph{Never} denotes no refresh after the initial context; $\mathrm{TR}=1$ s. Values and error-bar portions below zero are clipped.}
\label{fig:hrf_all_models_appendix}
\end{figure}
\clearpage
\endgroup

\end{document}